\documentclass[10pt,aps,twocolumn,showpacs,amssymb,prb]{revtex4-1}

\usepackage{graphicx}
\usepackage{amssymb}
\usepackage{amsmath}
\usepackage{xspace}
\usepackage{color}
\usepackage{braket}

\begin{document}

\title{Comparative density-matrix renormalization group study 
of symmetry-protected  topological phases in spin-1 chain 
and Bose-Hubbard models}

\author{Satoshi Ejima}
\author{Holger Fehske}
\affiliation{Institut f\"ur Physik, Ernst-Moritz-Arndt-Universit\"at
Greifswald, 17487 Greifswald, Germany}

\begin{abstract}
We reexamine the one-dimensional spin-1 $XXZ$ model with on-site uniaxial single-ion 
anisotropy as to the appearance and characterization of the
symmetry-protected topological Haldane phase.  
By means of large-scale density-matrix renormalization group (DMRG) calculations the central charge 
can be  determined  numerically via the von Neumann entropy, from which the ground-sate phase 
diagram of the model can be derived  with high precision. The nontrivial gapped Haldane phase 
shows up in between the trivial gapped even Haldane and N{\'e}el phases, appearing at large single-ion and 
spin--exchange interaction anisotropies, respectively. 
We furthermore carve out a characteristic degeneracy of the lowest entanglement level
in the topological Haldane phase, which is determined using a
conventional finite-system DMRG technique with both periodic and open
boundary conditions. Defining the spin and neutral gaps in analogy to 
the single-particle and neutral gaps in the intimately connected 
extended Bose-Hubbard model, we show that the excitation gaps in the
spin model qualitatively behave just as for the bosonic system. 
We finally compute the dynamical spin structure factor in the three
different gapped phases and find significant differences in the
intensity maximum which  might be used to distinguish these phases experimentally. 
\end{abstract}


\date{\today}
\maketitle

\section{Introduction}
One-dimensional (1D) quantum spin systems have received continued attention as 
paradigms for strongly correlated systems, because miscellaneous---and even  
exotic---phases can be realized within simple model Hamiltonians. 
By way of example the exactly solvable spin-1/2 antiferromagnetic (AFM) 
Heisenberg chain is known to be gapless, while for integer spin a gap
exists between the ground state and the first excited state,
as conjectured first by Haldane.\cite{Ha83}  
Especially for the spin-1 chain, the Haldane gap  was confirmed 
experimentally,\cite{BMAHGH86,RVRERS87} and the dynamical spin structure factor has been 
observed by inelastic neutron scattering, e.g., 
on Ni(C$_2$H$_8$N$_2$)$_2$NO$_2$ClO$_4$.\cite{MBRSE92} 
Affleck, Lieb, Kennedy, and Tasaki (AKLT) proposed a exactly solvable
model that offers valuable clues to the physics of the spin-1 Heisenberg
chain.\cite{AKLT87}  The so-called AKLT state [cf. Fig. 1 (a) below]
successfully describes the ground state of the Haldane phase.\cite{KT92}  
Also for the the spin-1 $XXZ$ model, the ground-state phase diagram 
has been determined---even if a single-ion anisotropy is
added~\cite{CHS03}---e.g., by the Lanczos exact diagonalization (ED) 
technique based on the level spectroscopy method.\cite{No95}

Currently, quantum integer-spin chains have attracted extraordinary interest 
from a topological point of view. The gapped ground states 
in the Haldane phase can be classified by the projective representations
of the underlying symmetry group.\cite{GW09,PTBO10} The odd Haldane (OH)
phase in odd-integer spin chains with two half-integer edge spins is a 
symmetry-protected topological (SPT) phase, because the odd-$S$ AKLT state
cannot be adiabatically connected to another trivial state
without undergoing a phase transition. On the other hand, the even
Haldane (EH) state in the even-integer spin systems with integer 
edge spins\cite{TONSNK11,OTNSNK11,OTNSNK11b,Tz12} is a trivial state, 
since the even-$S$ AKLT state is adiabatically connected to 
a trivial state without a bulk phase transition.\cite{PBTO12,KZMBP13} 

Interestingly, a hidden SPT phase analogous to the OH phase
was discovered in the extended Bose-Hubbard model (EBHM) with longer-range
repulsions.\cite{DBA06} This Haldane insulator (HI) phase, embedded between 
the Mott insulator (MI) and the density wave (DW) phases in the intermediate
coupling regime, exhibits the characteristic degeneracy of the
entanglement spectrum in the Haldane phase.\cite{ELF14}  
The excitation gaps at the quantum phase transition lines depend on 
their universality classes.\cite{DBA06,BDGA08} 
Beyond that, the dynamical density structure factor 
$S_{\rm EBHM}(k,\omega)$ significantly differs in the MI, DW, and HI 
states.\cite{ELF14}

On the basis of our recent EBHM study,\cite{ELF14} 
in the present work, we investigate the topological properties of the
odd Haldane phase in the anisotropic spin-1 $XXZ$ chain  
which, as we will show, can be taken as an effective model for the EBHM.
Using the density matrix renormalization group (DMRG) 
technique,\cite{Wh92,Wh93,JF07} first we determine the phase boundaries
by exploiting the central charge. In order to confirm the closing of the 
excitation gap at the trivial-nontrivial phase transition points, we simulate both the 
spin and neutral gaps.  We furthermore demonstrate the degeneracy of 
entanglement levels in the OH phase with both periodic (P) and 
open (O) boundary conditions (BC) [for the anisotropic spin-1 $XXZ$ chain it is well known 
how the edge spins should be treated in the latter case]. 
In order to experimentally detect the topological HI phase
in the EBHM, various dynamical quantities have been proposed.\cite{DBA06,Da13,ELF14}
Here we will examine the dynamical spin structure factor $S^{zz}(k,\omega)$  
for the spin-1 model by means of the dynamical DMRG
(DDMRG) technique.\cite{Je02b} We will demonstrate that 
the intensity maximum in $S^{zz}(k,\omega)$ features a gapped dispersion 
in the non-trivial Haldane phase as obtained for $S_{\rm EBHM}(k,\omega)$ in the EBHM.
Since this quantity is directly accessible by inelastic neutron scattering, significant differences 
in $S^{zz}(k,\omega)$  could be used to detect the various gapped phases. 

This paper is organized as follows. In the next section we establish 
the anisotropic spin-1 $XXZ$ model and the corresponding 
EBHM. The physical quantities of interests are introduced in Sec.~\ref{physical-quantities}. 
Large-scale (D)DMRG results for the anisotropic spin-1 $XXZ$ chain
will be presented and discussed in Sec.~\ref{DMRG-results}.
Section~\ref{summary} contains a brief summary and our main conclusions.

\section{Model Hamiltonians}
\label{model}
In this section we introduce the anisotropic spin-1 $XXZ$
model and get back to its established ground-state phase properties. 
We then define the extended Bose-Hubbard model and 
point out  the correspondences with an effective spin-1 $XXZ$ model.

\subsection{Spin-1 $XXZ$ model with single-ion anisotropy}

The Hamiltonian of the 1D spin-1 $XXZ$ model with on-site anisotropy is given by 
\begin{eqnarray}
 \hat{{\cal H}}&=&
   \sum_j [ J ( \hat{S}_j^x \hat{S}_{j+1}^{x}+\hat{S}_j^y \hat{S}_{j+1}^{y} )
           +J_z \hat{S}_{j}^{z} \hat{S}_{j+1}^z ]
 \nonumber \\      
 && +D\sum_j (\hat{S}_j^z)^2\ , 
\label{hamil}
\end{eqnarray}
where \mbox{\boldmath $\hat{S}$}$_j$ denotes a spin-$1$ operator. The parameter
$D$ represents the uniaxial single-ion anisotropy. 
The ground-state phase diagram of the model \eqref{hamil} exhibits various gapful 
and gapless phases, namely,  following the conventional notations, 
the Haldane phase, the large-$D$ phase, two $XY$ phases, 
the ferromagnetic phase, and the N{\'e}el phase.\cite{Sc86,NR89,CHS03} 
According to this different types of phase transitions occur between these phases: 
(i) A gapful-gapful Gaussian phase transition takes place 
between the large-$D$ phase and the Haldane phase with the central
charge $c=1$, 
(ii) the Haldane-N{\'e}el transition appears to be of the Ising
universality class with $c=1/2$, and 
(iii) a gapless-gapful Berezinskii-Kosterlitz-Thouless (BKT) transition 
emerges between the $XY$ phase and the Haldane or large-$D$ phase. 
In what follows we restrict ourselves to the parameter region 
where $J_z>0$ and $D>0$. 
Following the notation by Kj\"all {\it et al.},\cite{KZMBP13} 
we use the termini EH, OH, and AFM phases 
instead of the large-$D$, Haldane, and N{\'e}el phases, respectively.
The lattice inversion symmetry, which protects the SPT state of the 
Haldane phase, can be broken by adding a perturbation to the 
Hamiltonian (\ref{hamil}):
\begin{eqnarray}
 \delta\hat{\cal H}=g\sum_j [&&\hat{S}_j^z(\hat{S}_j^x \hat{S}_{j+1}^x + \hat{S}_j^y \hat{S}_{j+1}^y) 
  \nonumber\\
                        &&-\hat{S}_{j+1}^z(\hat{S}_j^x \hat{S}_{j+1}^x + \hat{S}_j^y \hat{S}_{j+1}^y)
                        +{\rm H.c.}
                       ]\, .
\label{pert-term}
\end{eqnarray}
Any finite $g$ immediately lifts the characteristic degeneracy of the lowest entanglement level 
in the Haldane phase.\cite{PTBO10} As we will see later, thereby the EH-OH quantum phase
transition also disappears.

\subsection{Extended Bose-Hubbard model}
In 2006, Dalla Torre {\it et al.}\cite{DBA06} discovered the HI phase 
in the 1D extended Bose-Hubbard model with longer-range repulsions. 
The HI phase features the properties of the OH phase
in the spin-1 model (\ref{hamil}).
The EBHM Hamiltonian reads  
\begin{eqnarray}
\hat{{\cal H}}_{\rm EBHM}&=&
 -t\sum_j( \hat{b}_j^\dagger \hat{b}_{j+1}^{\phantom{\dagger}}
          +{\rm h.c.})
 +U\sum_j \hat{n}_j(\hat{n}_j-1)/2 \nonumber \\
&& +V\sum_j \hat{n}_j\hat{n}_{j+1},
\label{EBHM-hamil}
\end{eqnarray}
where $\hat{b}_j^{\dagger}$ ($\hat{b}_j^{}$)  creates (annihilates) 
a boson at lattice site $j$,  and  
$\hat{n}_j=\hat{b}_j^\dagger\hat{b}_j$ 
is the corresponding boson number operator.   
The nearest-neighbor boson transfer amplitude is given by $t$ and 
$U$ ($V$) parametrizes the on-site (nearest-neighbor) particle repulsion.
Assuming that the site occupation is restricted to $n_j=0$, 1, or 2, 
with $S_j^z=n_j-1$ for a mean boson filling factor $\rho=N/L=1$,
the system can be mapped onto an effective spin-1 Hamiltonian, 
\begin{eqnarray}
 \hat{\cal H}_{\rm EBHM}^{\rm eff}=\hat{\cal H}+\hat{\cal H}^{\prime}, 
\label{eff-hamil-BH}
\end{eqnarray}
with the replacements $J\to -t$, $J_z\to V$, and $D\to U/2$
in Eq.~\eqref{hamil}. $\hat{\cal H}^{\prime}$ contains further terms
which breaks the particle-hole symmetry of $\hat{\cal H}$  
[see Eq.~(A1) of Ref.~[\onlinecite{BDGA08}] for the explicit 
form of $\hat{\cal H}^{\prime}$].
The EBHM exhibits three insulating phases, where  the nontrivial HI phase 
appears in between the MI and DW phases 
for intermediate-couplings. The MI, HI, and DW phases of the EBHM  
correspond to the EH, OH, and AFM phases of the spin-1 model~(\ref{hamil}), 
respectively. 

\section{Physical quantities of interest}
\label{physical-quantities}
In this section we assort the quantities that can be used to characterize the different  
phases and phase transitions in the spin-1 model (\ref{hamil}) and accordingly in the EBHM. 
We furthermore  explain how the quantities can be simulated using the DMRG technique. 

\subsection{Entanglement spectrum, von Neumann entropy, and central charge}
\begin{figure}[tb]
 \begin{center}
 \includegraphics[clip,width=0.9\columnwidth]{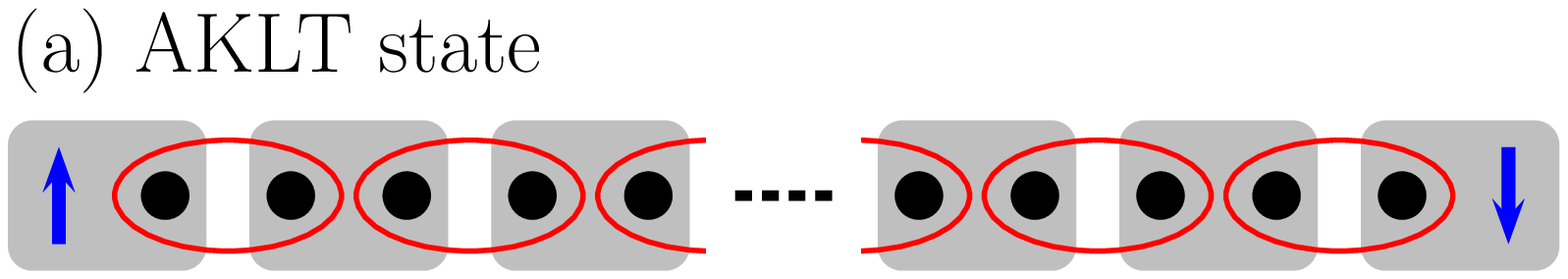}\\[0.2cm]
  \includegraphics[clip,width=0.9\columnwidth]{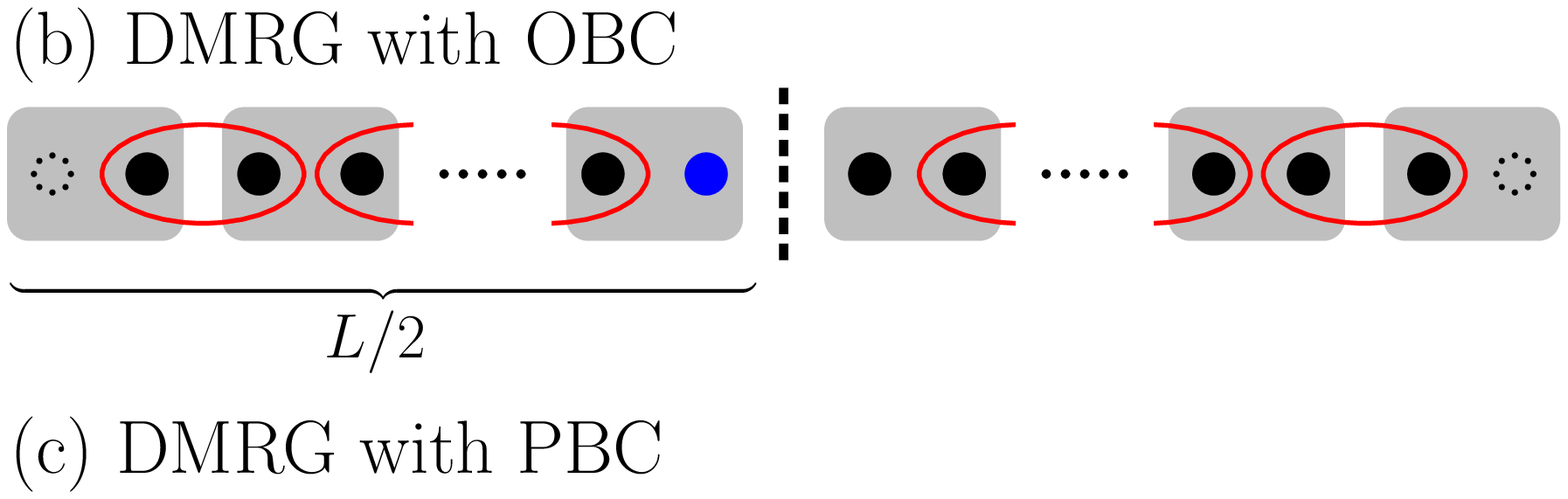}
  \includegraphics[clip,width=0.5\columnwidth,angle=-90]{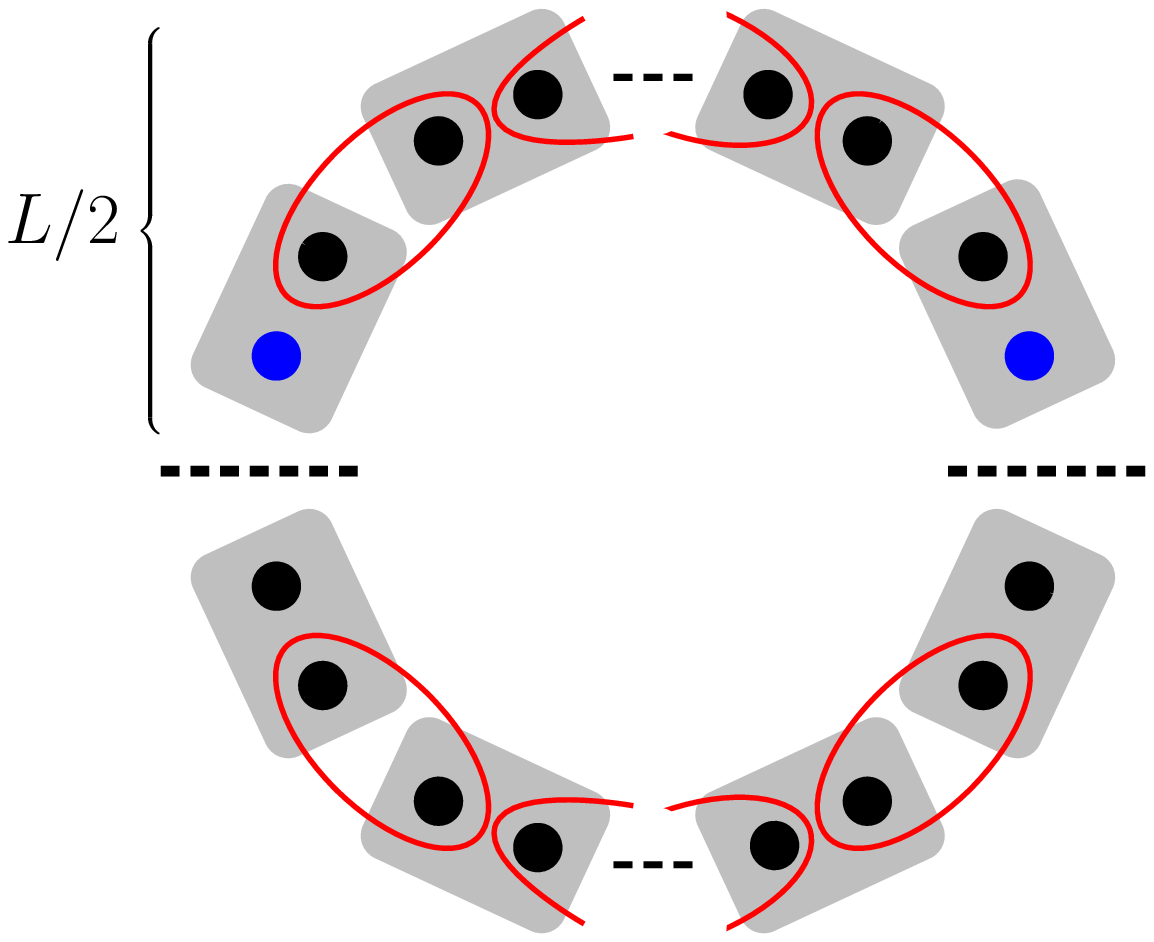}
 \end{center}
 \caption{(Color online) (a) Valence bond picture of 
 the AKLT state in a spin-1 $XXZ$ chain. 
 Each of the two $S=1/2$ spins connected by an ellipse form a singlet 
 ($1/\sqrt{2}$)($\uparrow\downarrow-\downarrow\uparrow$). 
 The two free edge-spins cause the fourfold degeneracy of the ground-state energy. 
 (b) To simulate the AKLT state within OBC DMRG, the free edge spins have to be 
 excluded from the system (dashed circles).  
 (c) With PBC the AKLT state can be simulated without any changes,
 so that the lowest entanglement level exhibits a fourfold degeneracy
 according to the two edge spins.
 }
\label{Haldane-state}
\end{figure}

After Li and Haldane's proposal~\cite{LH08} to characterize topological
phases by the entanglement spectrum this has become one of the most
powerful tools to investigate the SPT state.
Dividing a system with $L$ sites into two subblocks and considering
the reduced density matrix $\rho_\ell=\mathrm{Tr}_{L-\ell}[\rho]$ 
of a subblock of arbitrarily length $\ell$, the entanglement spectrum 
$\xi_\alpha$ is obtained from the weights  $\lambda_\alpha$ of the 
reduced density matrix $\rho_\ell$ by 
\begin{equation}
\xi_\alpha=-2\ln\lambda_\alpha\,.
\end{equation}
The entanglement spectrum of a subblock with $\ell=L/2$ sites can be
obtained for OBC and PBC as sketched in
Figs.~\ref{Haldane-state}(b) and \ref{Haldane-state}(c), respectively.
Thereby the artificial edges give rise to the characteristic degeneracy of the lowest 
entanglement level in the nontrivial AKLT state 
[displayed in Fig.~\ref{Haldane-state}(a)], 
where the degree of degeneracy depends on the boundary conditions.
To determine  the entanglement spectrum in the OH phase with OBC, 
a well-known trick is in use: 
One simulates a system without free edge spins by replacing the edge sites with
$S=1/2$, as shown in Fig.~\ref{Haldane-state}(b). One then expects a doubly 
degenerate lowest entanglement level in the OH phase. 
For PBC, on the other hand, for the same finite system, 
a fourfold degeneracy is expected due to two free $S=1/2$ spins 
[see Fig.~\ref{Haldane-state}(c)],  
just as for the HI phase in the EBHM.\cite{ELF14} 

The entanglement analysis provides also valuable information 
about the criticality of the system.
Adding up the $\lambda_\alpha$ during the simulation, we have direct access 
to the von Neumann entropy $S_L(\ell)=-\mathrm{Tr}_\ell[\rho_\ell\ln\rho_\ell]$.
From conformal field theory,\cite{CC04} it follows that in the case of a periodic
system the von Neumann entropy takes the form 
\begin{eqnarray}
 S_L(\ell)=\frac{c}{3}\ln\left[\frac{L}{\pi}\sin\left(\frac{\pi\ell}{L}\right)\right]
           +s_1\, ,
\end{eqnarray}
where $s_1$ is a non-universal constant. 
Since the most precise data of $S_L(\ell)$ are obtained when the length $\ell$ of the 
subblock equals half the system size $L$,  the relation~\cite{Ni11}  
\begin{eqnarray}
 c^\ast(L) \equiv \frac{3[S_L(L/2-1)-S_L(L/2)]}{\ln[\cos(\pi/L)]}
\label{cstar}
\end{eqnarray}
is much better suited for determining the central charge than directly 
using the above expression for $S_L(\ell)$.

For the EBHM the phase boundaries can be assigned very effectively using the
(numerically determined) central charge $c^\ast$, because the system becomes critical 
only at the MI-HI (HI-DW) transition points where $c=1$ ($c=1/2$), 
and there $c^\ast$ shows pronounced peaks.\cite{ELF14} 
Hence we adopt this method for the spin-1 model~\eqref{hamil} as well 
to pinpoint the  EH-OH and OH-AFM transition points.

\subsection{Excitation gaps}
\label{sec:gap}
Monitoring various excitation gaps for the EBHM, significant features have been found 
at the MI-HI and HI-DW transition points.\cite{DBA06,BDGA08}
For example, the single-particle gap,
\begin{equation}
\Delta_c=E_0^{\rm EBHM}(N+1)+E_0^{\rm EBHM}(N-1)-2E_0^{\rm EBHM}(N)\,,
\label{spg}
\end{equation}
is finite in all three insulating phases, except for the MI-HI 
transition point. 
By contrast, the neutral gap, 
\begin{equation}
\Delta_n=E_1^{\rm EBHM}(N)-E_0^{\rm EBHM}(N)\,,
\label{ng}
\end{equation}
closes at both the MI-HI and HI-DW transitions.  In Eqs.~\eqref{spg} 
and~\eqref{ng}, $E_0^{\rm EBHM}(N)$ and $E_1^{\rm EBHM}(N)$ denote 
the energies of the ground state and first excited state 
of the $N$-particle system for the EBHM, respectively. 

Since adding (removing) a particle in the EBHM corresponds to 
raising (lowering) the spin $S^z$ projection in a pseudospin model, 
we consider for the spin-1 $XXZ$ model the spin gap,
\begin{eqnarray}
 \Delta_s=E_0^{XXZ}(1)-E_0^{XXZ}(0)\,,
\end{eqnarray}
which likewise might be finite in all three phases, except for the EH-OH
transition point. As for the EBHM, the neutral gap in the spin-1 
model (\ref{hamil}) can be defined as
\begin{eqnarray}
 \Delta_n=E_1^{XXZ}(0)-E_0^{XXZ}(0),
\end{eqnarray}
where $E_0^{XXZ}(M)$ and $E_1^{XXZ}(M)$ denote the ground-state 
and first excited energies within the subspace $M=\sum_jS_j^z$,  respectively.
By analogy to the behavior of the neutral gap in the EBHM, 
$\Delta_n$ should vanish at the EH-OH and OH-AFM transition points 
for the spin-1 chain model.

\subsection{Dynamical spin structure factor}
Simulating the dynamical spin structure factor by DDMRG is of particular importance
since it might be directly compared with inelastic neutron scattering
experiments, e.g., on Ni(C$_2$H$_8$N$_2$)$_2$NO$_2$ClO$_4$.\cite{MBRSE92} 
Its $zz$-component is defined by 
\begin{eqnarray}
  S^{zz}(k,\omega)=\sum_n |\langle \psi_n|\hat{S}_k^z|\psi_0\rangle|^2
             \delta(\omega-\omega_n)\,, 
\end{eqnarray}
where $|\psi_0\rangle$ and $|\psi_n\rangle$ denote the ground state and
$n$th excited state, respectively.  The corresponding excitation energy
is $\omega_n=E_n-E_0$. For $D=0$, i.e., for the isotropic Heisenberg or 
$XXZ$ fix points of~(\ref{hamil}), 
$S^{zz}(k,\omega)$ was extensively studied by ED~\cite{Ta94b} 
and time-dependent DMRG~\cite{WA08} techniques. 
That is, the behavior of $S^{zz}(k,\omega)$ in the Haldane
phase is well known, albeit numerical results for the EH and AFM states are rare. 
Taking into account the relation $S_j^z=n_j-\rho$ for the pseudospin 
in the effective model $\hat{\cal H}_{\rm EBHM}^{\rm eff}$, one expects 
that the spin structure factor $S^{zz}(k,\omega)$ corresponds to the dynamical
density structure factor $S_{\rm EBHM}(k,\omega)$ in the EBHM, which exhibits 
different behavior  in the three insulating phases.\cite{ELF14}

\section{Numerical results}
\label{DMRG-results}
In this section we present our numerical (D)DMRG results for the spin-1 
$XXZ$ model with and without single-ion anisotropy.  
We first determine the phase boundaries and then analyze  the behavior of the 
excitation gaps at the transitions between the nontrivial and trivial phases.
Furthermore, we discuss the entanglement spectra of an odd Haldane phase. 
Finally, we simulate the dynamic spin structure factor and compare it with the 
dynamical density response in the EBHM.

In the numerics we keep up to $m=3200$ density-matrix states for the 
static DMRG runs, so that the discarded weight is typically smaller than 
$1\times10^{-10}$. For the DDMRG simulations we take $m=800$,  
examining the ground state along the first five DMRG sweeps, and then 
use $m=400$ states computing dynamical properties.

\subsection{Phase boundaries}

\subsubsection{OH-AFM transition}

\begin{figure}[tb]
 \begin{center}
  \includegraphics[clip,width=\columnwidth]{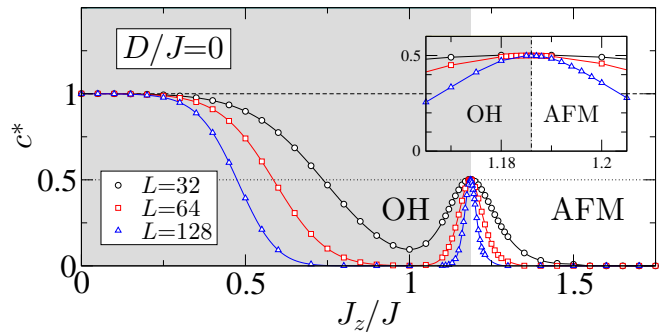}
 \end{center}
 \caption{(Color online)
 Central charge $c^\ast(L)$ as obtained by DMRG for 
 the spin-1 $XXZ$ model with $D=0$ and PBC.
  The OH-AFM transition can be assigned to
 $J_z/J=1.186\pm0.001$ with $c=1/2$, where a pronounced peak appears
 [see also the magnifying inset which shows  $c^\ast(L)$ close to the 
 transition point].
 }
\label{cstar-D0}
\end{figure}

Let us first discuss the spin-1 model (\ref{hamil}) with $D=0$.
In this case it is known that a BKT transition occurs 
at $J_z=0$ between the $XY$ and OH phases.\cite{KNO96} 
At $J_z>0$, only an OH-AFM transition takes place,
where $c=1/2$ is expected.

Figure~\ref{cstar-D0} shows the central charge
$c^\ast(L)$, computed from Eq.~\eqref{cstar}.
If $J_{z}/J$ is raised at fixed system size, the maximum in $c^\ast(L)$ sharpens at the OH-AFM transition point 
$J_{z,{\rm c1}}/J$, and we deduce $c^\ast\simeq0.5$. 
The other critical point $J_{z,{\rm c2}}/J\simeq 1.185$
with $c^\ast\simeq 0.503$ approximates  the recent infinite-system DMRG result
$J_{z,{\rm c2}}/J=1.186\pm0.002$\cite{KZMBP13,LLYST14} very well already for $L=32$.
The agreement becomes perfect if we increase the system size: 
$J_{z,{\rm c2}}/J\simeq 1.186$ with $c^\ast\simeq0.500$ for $L=128$.
Note that $c^\ast(L)$ stays equal to one in a relatively wide region (from $J_{z}/J=0$ to $J_{z}/J\simeq 0.3$ for $L=128$), indicating the BKT transition between the $XY$ and the OH phases at $J_{z}/J=0$ with $c=1$.

\begin{figure}[tb]
 \begin{center}
  \includegraphics[clip,width=\columnwidth]{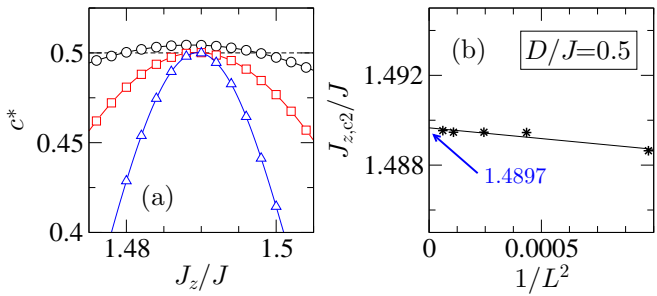}
 \end{center}
 \caption{(Color online)
 $c^\ast(L)$ for $D=0.5$ near the OH-AFM transition. Panel (b) shows that
 the Ising transition point $J_{z,c2}/J(L)$ obtained from $c^\ast(L)$ 
 can be linearly extrapolated to the thermodynamic limit.  
 }
\label{cstar-D0.5}
\end{figure}

To relate our numerical results to previous ones  
we include an on-site anisotropy $D$ and compute $c^\ast(L)$
in the vicinity of the OH-AFM transition. For $D/J=0.5$ the central charge 
$c^\ast(L)$ at fixed system size $L$ develops again a pronounced maximum 
at the OH-AFM transition point [see Fig.~\ref{cstar-D0.5}(a)]. 
The  deduced transition point $J_{z,\, c1}/J(L)$ is readily extrapolated 
to the thermodynamic limit [Fig.~\ref{cstar-D0.5}(b)], 
yielding $J_{z,\,c2}/J\simeq1.4897$, which is in reasonable agreement 
with the ED result $J_{z,\,c2}/J\simeq1.536$
obtained from systems with $L$ up to 16\cite{CHS03} and confirms  
recent DMRG data  $J_{z,\,c2}/J=1.4905\pm0.0015$.\cite{UNK08}

\subsubsection{EH-OH transition}
\begin{figure}[tb]
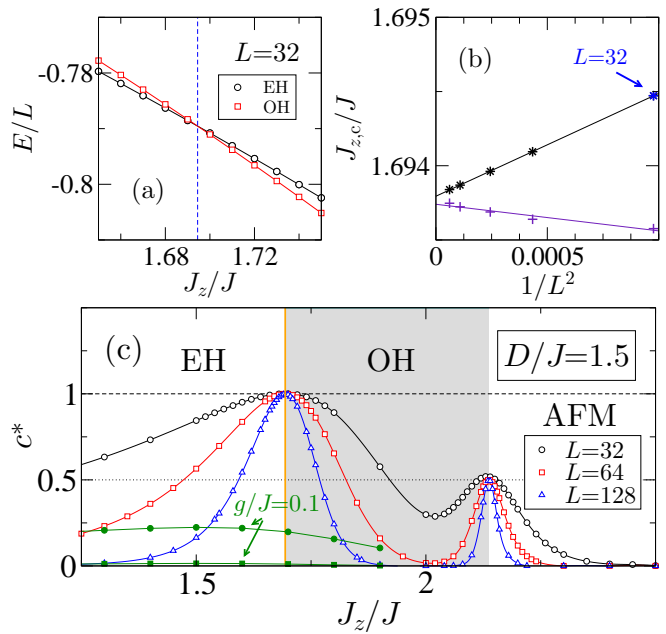

 \begin{center}
  \includegraphics[clip,width=\columnwidth]{fig4ab.eps}
  \includegraphics[clip,width=\columnwidth]{fig4c.eps}
 \end{center}
 \caption{(Color online) 
 (a): $J_z$-dependence of the two lowest energy eigenvalues 
 at $D/J=1.5$, using TBC and $L=32$. 
 Obviously  the energies of the OH state (squares) and the EH state 
 (circles) cross at the EH-OH transition point (dashed line). 
 (b): Critical points $J_{z,\, {\rm c1}}/J$ obtained by the level
 spectroscopy technique (stars) [via the central charge (pluses)] 
 as obtained in panel~(a) [(c)] versus the inverse of the 
 system size squared at $D/J=1.5$ for $L$ up to~128. 
 (c): Central charge $c^\ast$ of the 1D spin-1 $XXZ$ model~(\ref{hamil}) with $D/J=1.5$, 
 indicating the EH-OH (OH-AFM) transition point with $c=1$ ($c=1/2$). 
 The solid line denotes the EH-OH transition extracted from panels (a) and (b), 
 which is in accordance with the position of the maximum in $c^\ast(L)$. 
 Turning on a perturbation
 $\delta\hat{\cal H}$ that breaks the lattice-inversion symmetry, 
 the central charge $c^\ast(L)$ (filled symbols) 
 becomes zero for large enough system sizes ($L\geq64$).
 }
\label{twist-cstar-D1.5}
\end{figure}

We now turn to the case $D>0$.  
In previous works~\cite{KNO96,KN97a,KN97b,CHS03} a Gaussian transition between 
the EH and OH phases has been found by employing the level spectroscopy technique
to ED results obtained for small systems.    
Applying the twisted boundary conditions (TBC), 
$\hat{S}_{L+1}^x=-\hat{S}_1^x$, $\hat{S}_{L+1}^y=-\hat{S}_1^y$, and 
$\hat{S}_{L+1}^z=\hat{S}_1^z$ within DMRG, the two lowest energy 
levels can be simulated accurately for much larger
system sizes than accessible to ED. Figure~\ref{twist-cstar-D1.5}(a)
demonstrates that the two lowest energies assigned  to 
the EH and OH states  cross at $J_{z,\,c1}/J\simeq1.6945$ by 
increasing $J_z/J$ at fixed $D/J=1.5$ for $L=32$ 
(and TBC). The critical points $J_{z,\,c1}/J(L)$ 
can be systematically extrapolated to the thermodynamic limit by a linear fit, 
as indicated in Fig.~\ref{twist-cstar-D1.5}(b).
For $L\to \infty$ we obtain $J_{z,\,c1}/J\simeq1.6938$.

Alternatively, the EH-OH transition points can be extracted from the  
central charge $c^\ast(L)$ if compared with the field theoretical
prediction $c=1$. This is demonstrated in Fig.~\ref{twist-cstar-D1.5}(c).
Here the maxima of $c^\ast(L)$ can also be extrapolated to the
thermodynamic limit [see Fig.~\ref{twist-cstar-D1.5}(b)], where 
transition point is in excellent accord with the ones 
via level spectroscopy in Figs.~\ref{twist-cstar-D1.5}(a) 
and \ref{twist-cstar-D1.5}(b).
We note that also the OH-AFM transition can be reliably determined from the 
peak at $J_{z,\,c2}/J\simeq2.138$.

\subsection{Characterization of the topological phase}
In the following we analyze the signatures of the topological
OH phase and of the transition between the  trivial and nontrivial topological
states for the model~\eqref{hamil} in close analogy to the EBHM.\cite{ELF14} 
To this end, we simulate the excitation gaps and the entanglement spectra.

\subsubsection{Excitation gaps}
\begin{figure}[tb]
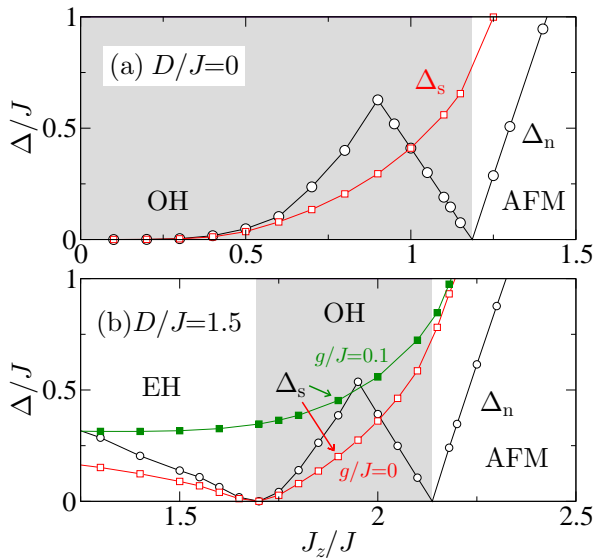

 \begin{center}
  \includegraphics[clip,width=0.9\columnwidth]{fig5a.eps}
  \includegraphics[clip,width=0.9\columnwidth]{fig5b.eps}
 \end{center}
 \caption{(Color online) 
 Extrapolated data for the spin gap $\Delta_{\rm s}$
 (squares) and neutral gap $\Delta_{\rm n}$ (open circles) 
 as a function of $J_z/J$ for   $D/J=0$ [panel (a)] and  $D/J=1.5$ [panel (b)].
 The filled squares in panel (b) give $\Delta_{\rm s}$ with
 a finite inversion-symmetry breaking perturbation 
 $g/J=0.1$ [see Eq.~\eqref{pert-term}].
 }
\label{gaps}
\end{figure}

So far the excitation gaps of~\eqref{hamil} have been studied
mostly at the isotropic Heisenberg point with respect to  
the magnitude of the Haldane gap. 
At the trivial-nontrivial phase transition points the excitation gaps
should close, as demonstrated, e.g., for the EBHM.\cite{DBA06,BDGA08}
Here we compute the spin and neutral excitation gaps as defined in Sec.~\ref{sec:gap} 
instead of calculating the simple first excitation gap. Thereby, we adopt PBC instead of OBC within DMRG, 
avoiding the use of edge spins, which have to be adapted according to the considered parameter
region.

Figure~\ref{gaps}(a) shows first the excitation gaps at $D=0$. 
Upon increasing $J_z/J$, the gaps open exponentially, reflecting
the BKT transition at $J_z/J=0$. $\Delta_{\rm n}$ and $\Delta_{\rm s}$
cross each other exactly at the Heisenberg point, $J_z/J=1$, 
where $\Delta_{\rm n}(L)=\Delta_{\rm s}(L)$ 
(see the discussion about the system-size dependence of the excitation gaps
and the magnitude of the Haldane gap for the spin-1 Heisenberg model
in the Appendix).
At the OH-AFM transition ($J_{z,\,c}/J\simeq 1.186$), $\Delta_n$
closes linearly  because the transition belongs to the Ising universality class, while 
$\Delta_s$ remains finite. 

For $D/J=1.5$ [see Fig.~\ref{gaps}(b)], the EH-OH transition occurs at
$J_{z,\,c1}/J\sim1.6938$, where both spin and neutral gaps vanish.
Increasing $J_z/J$ above  $J_{z,\,c1}/J$, only $\Delta_n$ closes
at the Ising transition point $J_{z,\,c2}/J$, just as in the case of 
$D/J=0$ [compare Figs.~\ref{gaps}(a) and \ref{gaps}(b)]. 
If we turn on the perturbation $\delta\hat{\cal H}$ [see Eq.~\eqref{pert-term}], which 
breaks the lattice-inversion symmetry explicitly, the EH-OH transition disappears, 
so that $\Delta_s$ stays finite for $g/J=0.1$,  as shown in Fig.~\ref{gaps}(b).
Thereby, owing to the loss of the criticality at the EH-OH transition, 
$c^\ast(L)$ converges to zero for large enough $L$, as demonstrated
in Fig.~\ref{twist-cstar-D1.5}(c).

Comparing the behavior of the excitation gaps with those of the EBHM,\cite{DBA06,BDGA08} 
one sees that the spin (neutral) gap in the spin-1 model \eqref{hamil} takes the role of the
single-particle (neutral) gap in the EBHM.

\subsubsection{Entanglement spectra}
\begin{figure}[tb]
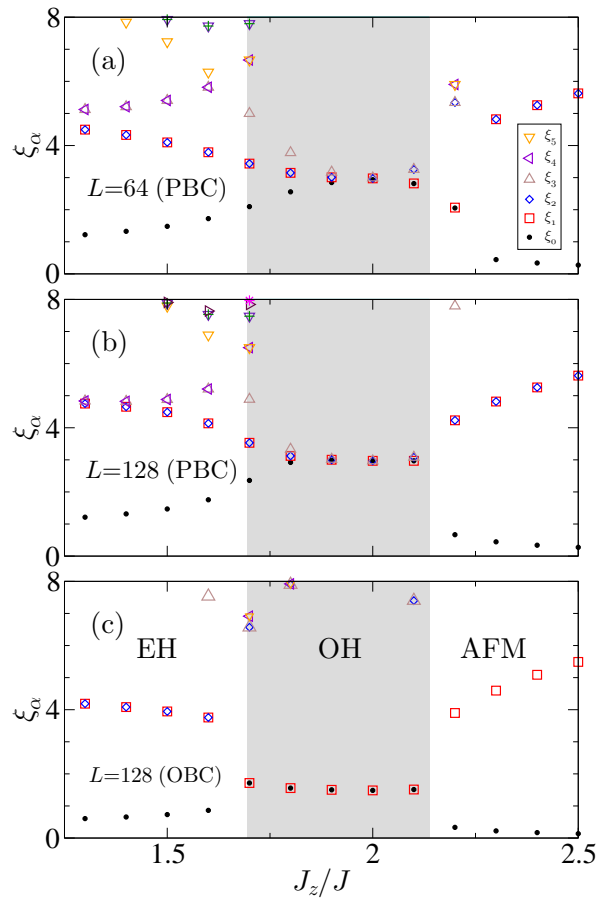

 \begin{center}
  \includegraphics[clip,width=0.9\columnwidth]{fig6a.eps}
  \includegraphics[clip,width=0.9\columnwidth]{fig6b.eps}
  \includegraphics[clip,width=0.9\columnwidth]{fig6c.eps}
  \end{center}
 \caption{(Color online) Entanglement spectrum $\xi_{\alpha}$ of the
 1D spin-1 $XXZ$ model~(\ref{hamil}) with $D=1.5$. The fourfold degeneracy 
 of the lowest entanglement level can be taken as an indication of 
 a nontrivial Haldane state in the case of DMRG simulations with PBC. 
 As the system size increases the degeneracy  appears in the  
 whole HI phase; compare data for $L=64$ [(a)] with those 
 for $128$ [(b)]. Using OBC and taking spins ($S=1/2$) at the edges into
 account, an almost perfect double degeneracy is obtained for the OH phase 
 even for small systems with  $L=128$, as demonstrated in (c).
 }
\label{ES-D1.5}
\end{figure}

Let us now analyze the entanglement properties of the topological states
for intermediate single-ion anisotropy ($D/J=1.5$),  
where both the EH-OH and OH-AFM transitions exist. 
Here Pollmann {\it et al.}\cite{PTBO10} showed that 
the SPT state in the OH phase has a twofold degenerate 
lowest entanglement level for the quantum spin chain model.
The infinite-time evolving block decimation procedure used by those authors, 
gives the entanglement spectra data directly in the thermodynamic limit. 
In the following we show that when simulating the model~\eqref{hamil} 
for a finite system by conventional DMRG,  
this characteristic degeneracy of the OH phase can also be obtained, but 
the degree of the degeneracy depends on the boundary conditions.

Figure~\ref{ES-D1.5} presents the entanglement spectrum $\xi_\alpha$ for  the 
anisotropic spin-1 $XXZ$ model with $D/J=1.5$. For a small system  ($L=64$)
with PBC [Fig.~\ref{ES-D1.5}(a)] the lowest entanglement level is fourfold
degenerate only deep inside the OH phase.
This reflects the possession of the two edges for the subblock $L/2$. 
Increasing the system size this degeneracy is observed  for a larger region of the OH phase, 
as demonstrated by Fig.~\ref{ES-D1.5}(b) for $L=128$, but close to the EH-OH transition point 
the lowest entanglement level is still non-degenerate. 
To overcome this drawback we apply OBC with half-spin edges 
in the OH phase [cf. Fig.~\ref{Haldane-state}(b)]. The same procedure has been used 
to  estimate the magnitude of the Haldane gap at the isotropic Heisenberg point. 
Figure~\ref{ES-D1.5}(c) gives $\xi_\alpha$ for $L=128$ and OBC, pointing out 
the twofold degeneracy of the lowest level in the nontrivial phase and its non-degeneracy  
anywhere else. The degeneracy is clearly caused by the single 
edge spin of subblock $L/2$.

\begin{figure}[t]
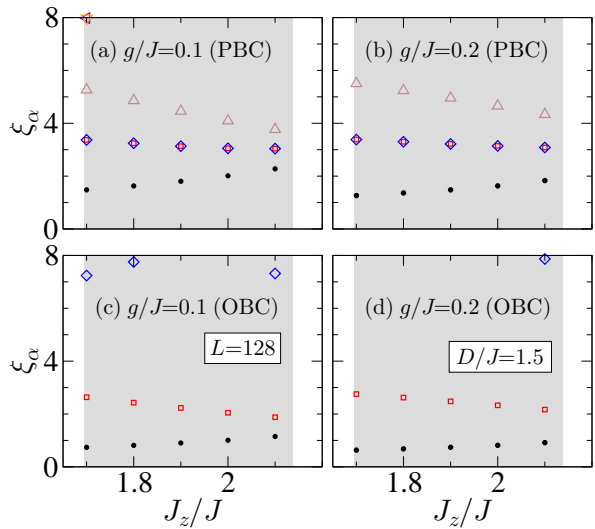

 \begin{center}
  \includegraphics[clip,width=0.9\columnwidth]{fig7ab.eps}
  \includegraphics[clip,width=0.9\columnwidth]{fig7cd.eps}
  \end{center}
 \caption{(Color online) Entanglement spectrum $\xi_{\alpha}$ if an 
 inversion-symmetry-breaking term is added to the spin-1 chain (\ref{hamil}) with $D=1.5$, 
 where $g/J=0.1$ (left panels) and $0.2$ (right panels). Data obtained by  
  DMRG with PBC (upper panels) and OBC with half-spin edges (lower panels).
 }
\label{ES-D1.5-perturb}
\end{figure}

Recently it has been demonstrated for quantum spin chains\cite{PTBO10} 
and the EBHM\cite{ELF14} that the degeneracy of the lowest entanglement 
level in the OH phase might be lifted by turning on an 
inversion-symmetry-breaking term, such as ~\eqref{pert-term}.
Figure~\ref{ES-D1.5-perturb}(a) [Figure~\ref{ES-D1.5-perturb}(c)]
exemplifies that the fourfold [twofold] degeneracy with PBC [OBC] 
indeed dissolves for any finite $g$. 
Thereby, the gap between the lowest levels becomes larger as $g$ increases, 
see Fig.~\ref{ES-D1.5-perturb}(b) [Fig.~\ref{ES-D1.5-perturb}(d)] 
for PBC [OBC]. Obviously inversion symmetry protects the Haldane phase.

\subsubsection{Multicritical point and EH-AFM transition}

\begin{figure}[t]
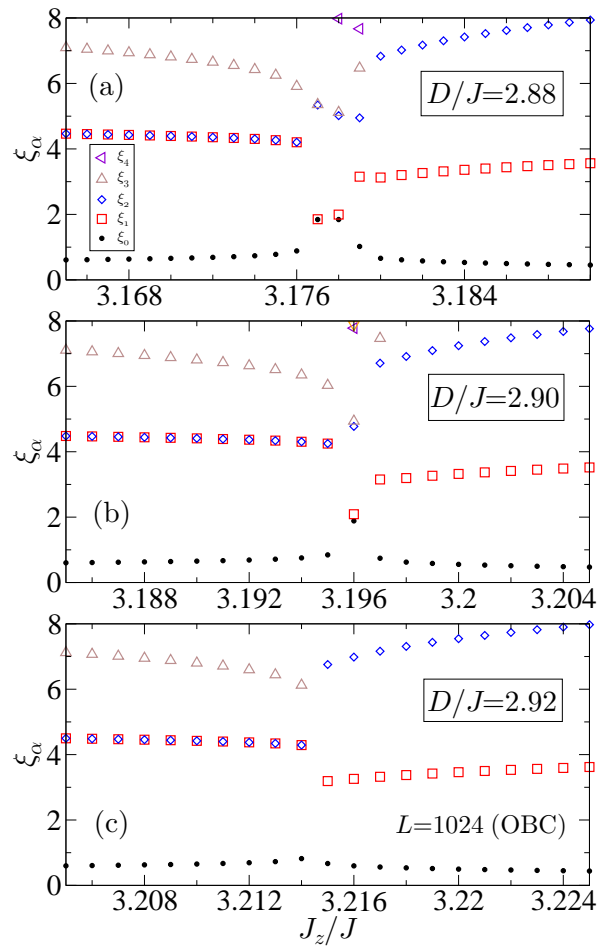

 \begin{center}
  \includegraphics[clip,width=0.9\columnwidth]{fig8a.eps}
  \includegraphics[clip,width=0.9\columnwidth]{fig8b.eps}
  \includegraphics[clip,width=0.9\columnwidth]{fig8c.eps}
  \end{center}
 \caption{(Color online) Entanglement spectrum $\xi_{\alpha}$ 
 for $D/J=2.88$ (a), $2.90$ (b), and $2.92$ (c)
 close to the EH-OH-AFM multicritical point with $L=1024$ and OBC,
 showing the disappearance of the double degeneracy of the lowest
 entanglement levels.
 }
\label{Tricritical}
\end{figure}

Raising the ratios $D/J$ and $J_z/J$, the EH-OH and OH-AFM critical lines merge 
at the multicritical point  ($J_{z,{\rm mc}}/J$, $D_{\rm mc}/J$). Above
this point, i.e., for $J_z>J_{z,{\rm mc}}$ and $D>D_{\rm mc}$,  a direct EH-AFM transition is 
expected to occur, as pointed out by den Nijs and Rommeles.\cite{NR89}  This has been confirmed  
numerically by ED, yielding ($J_{z,{\rm mc}}$, $D_{\rm mc}$) $\approx$ (3.2, 2.9).\cite{CHS03} 
It is challenging to determine this multicritical point more precisely,
but the entanglement analysis outlined above seems to be a powerful tool.
Obviously, for fixed values $D>D_{z,{\rm mc}}$, the lowest 
entanglement level is non-degenerate for the whole parameter regime of
$J_z/J$ including the EH-AFM transition point,
while for $D<D_{\rm mc}$ the lowest entanglement level should be degenerate 
for a very narrow but finite parameter region of $J_z/J$. 
In fact, the double degeneracy
can still be observed at $D/J=2.88$ by large-scale DMRG simulations
with $L=1024$ and OBC [see Fig.~\ref{Tricritical}(a)].
If $D/J$ is increased slightly, the degeneracy is lifted for $L=1024$
[see Fig.~\ref{Tricritical}(b) for $D/J=2.9$], but from the results presented 
we cannot derive a definite conclusion about what happens for $L\to \infty$.
Figure~\ref{Tricritical}(c) indicates that degeneracy disappears
already at $D/J=2.92$. In this way the ED results regarding   
the existence of the multicritical point is corroborated by our more precise entanglement spectra 
analysis, yielding ($J_{z,{\rm mc}}/J$, $D_{\rm mc}/J$)=($3.196\pm0.02$, $2.90\pm0.02$).

\begin{figure}[tb]
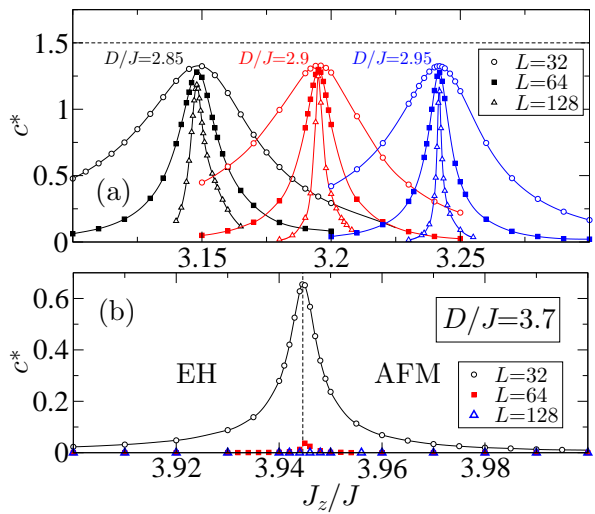

 \begin{center}
  \includegraphics[clip,width=0.9\columnwidth]{fig9a.eps}
  \includegraphics[clip,width=0.9\columnwidth]{fig9b.eps}
 \end{center}
 \caption{(Color online) 
 Central charge $c^\ast$ of the 1D spin-1 $XXZ$ model~(\ref{hamil})
 near the multicritical point $D_{\rm mc}/J$ (a) and across the 
 first-order EH-AFM transition for $D/J=3.7$ (b). The dashed line 
 in panel (b) denotes the EH-AFM transition point $D/J\simeq 3.9446$
 according to Ref.~[\onlinecite{LLYST14}].
 }
\label{cstar-multi-EH-AFM}
\end{figure}

Certainly it is of great interest to look at the behavior of the 
central charge at the multicritical point. 
Here the central charge might be $c=1+1/2=3/2$ 
because BKT- and Ising-transition lines merge. 
Figure~\ref{cstar-multi-EH-AFM}(a) displays   
the numerically obtained central charge $c^\ast(L)$ for fixed 
values of $D/J$ in the vicinity of the multicritical point. 
We see that $c^\ast(L)$ is always smaller than 3/2 
and decreases with increasing system size $L$. 
Unfortunately, the system-size dependence of $c^\ast(L)$ is 
much stronger than those, e.g., at $D/J=1.5$, in Fig.~\ref{twist-cstar-D1.5};
so it turns out that even $L=128$ is not large enough to determine the value
of the central charge precisely. Maybe the use of the infinite-system 
DMRG~\cite{KZMBP13} can resolve this problem.

Increasing $D/J$ further, a quantum phase transition occurs
between EH and AFM phases. A  discontinuous  staggered magnetization 
suggests that this transition is of first order.\cite{CHS03} 
Quite recently, this was corroborated by analyzing the energy level 
crossing.\cite{LLYST14} The numerically determined central charge
$c^\ast(L)$ at $D/J=3.7$ yields a further signature of the 
first-order transition [see Fig.~\ref{cstar-multi-EH-AFM}(b)]. 
For small system sizes ($L=32$), $c^\ast$ shows a peak at $J_z/J\simeq3.945$, 
in accord with the EH-AFM transition point in Ref.~[\onlinecite{LLYST14}]. 
With increasing the system size $L$, $c^\ast$ decreases drastically and 
becomes already zero for $L=128$, 
which confirms the results of former studies.\cite{CHS03,LLYST14}

\subsection{Ground-state phase diagram}
\begin{figure}[t]
 \begin{center}
  \includegraphics[clip,width=0.9\columnwidth]{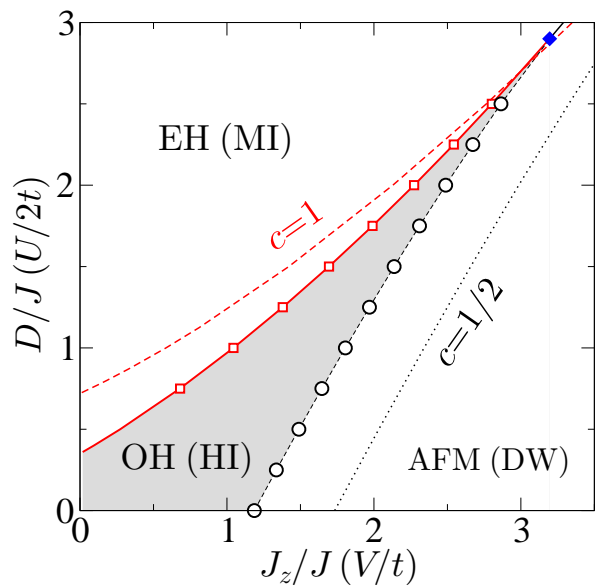}
 \end{center}
 \caption{(Color online) 
 DMRG ground-state phase diagram of the 1D spin-1 $XXZ$ model with single-ion anisotropy~\eqref{hamil}. 
 Shown are the even Haldane (EH), odd Haldane (OH), and 
 antiferromagnetic (AFM) phases. The EH-OH (squares) and OH-AFM
 (circles) transition points are determined from the central charge
 $c=1$ and $c=1/2$, respectively, which was extracted from the
 von Neumann entropy via Eq.~\eqref{cstar}.  
 The EH-OH transition line was confirmed by a careful finite-size
 scaling of the two low-lying energy levels with TBC.
 The filled diamond gives the EH-OH-AFM multicritical point
 determined from the entanglement analysis. Error bars are smaller than symbols.
 The dashed (dotted) line denotes the MI-HI (HI-DW) transition 
 in the EBHM with $n_b=2$ bosons per site (taken from 
 Ref.~[\onlinecite{ELF14}]).
 }
\label{pd}
\end{figure}

Figure~\ref{pd} displays the DMRG ground-state phase diagram of the
spin-1 $XXZ$ model with single-ion aniso\-tropy. The EH-OH and OH-AFM 
phase boundaries can been derived from central charge $c^\ast$, as explained above:  
Again we obtain a very good agreement with former ED and DMRG data.\cite{CHS03,UNK08}
Most notably, the nontrivial OH phase appears in between the trivial EH and AFM phases, 
just as the topological HI phase develops between the MI and DW phases in the EBHM. 
Therefore, we have included in Fig.~\ref{pd}, the phase boundaries of the MI-HI and HI-DW
transitions for the EBHM with $n_b=2$ (taken from Ref.~[\onlinecite{ELF14}]).  
Qualitatively, the phase diagram of the spin-1 model looks quite  similar to  those of the EBHM, 
except for the existence of the superfluid (SF) phase in the EBHM (not shown). 
Quantitatively, the topological phase of the EBHM captures a larger region in parameter space
than the OH phase however. This might be caused by the particle-hole symmetry-breaking term
$\hat{\cal H}^\prime$ in Eq.~\eqref{eff-hamil-BH}.

\subsection{Dynamical structure factor}

Let us finally discuss the spin dynamical properties of the spin-1 $XXZ$ model.
\begin{figure*}[t!]
 \includegraphics[clip,width=\textwidth]{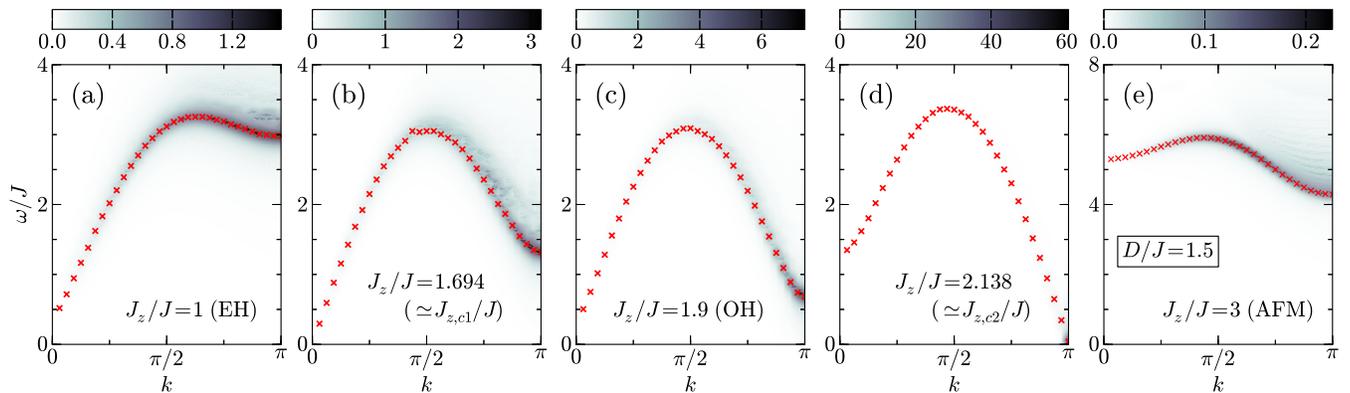}
 \caption{(Color online).
 Intensity plots of the dynamical structure factor $S^{zz}(k,\omega)$ 
 for (a) $J_z/J=1$, (b) $J_z/J=1.694\simeq J_{z,c1}/J$, (c) $J_z/J=1.9$,
 (d) $J_z/J=2.138\simeq J_{z,c2}/J$, and (e) $J_z/J=3$.
 Data are obtained by the DDMRG technique for $L=64$, using PBC 
 and a Lorenzian broadening $\eta=0.1t$.
 Crosses give the maximum value of $S^{zz}(k,\omega)$ at fixed momenta $k=2\pi j/L$
 with $j=1,\,\cdots,\, L/2$.
 }
 \label{Skw}
\end{figure*}
Figure~\ref{Skw} reveals our DDMRG results for $S^{zz}(k,\omega)$ 
obtained for the spin-1 model (with anisotropy $D/J=1.5$) inside the
three insulating phases, as well as at the  quantum phase transition points in between.  
In the EH phase, at $J_z/J=1$,  most of the spectral weight is concentrated  
in the momentum range $\pi/2<k<\pi$ [see Fig.~\ref{Skw}(a)]. 
The excitation gap appears at $k\approx0$. 
The dispersion of the maximum in
$S^{zz}(k,\omega)$ behaves cosine-like for small-to-intermediate
momenta, and is flattened close to the Brillouin zone boundary (above $k\ge3\pi/4$). 
With increasing $J_z/J$ the EH-OH transition occurs 
at $J_z/J=J_{z,c1}/J\simeq 1.694$, where the excitation gap closes 
at the momentum $k=0$, as shown in Fig.~\ref{Skw}(b).
Deep in the Haldane phase, the situation changes drastically 
[see Fig.~\ref{Skw}(c) for $J_z/J=1.9$]. 
Now the dispersion of the maximum in $S^{zz}(k,\omega)$ takes a
sine-like form. Again there are finite excitation gaps at $k=0$
(Haldane gap) and $\pi$. This resembles 
the behavior found at the isotropic Heisenberg point.\cite{Ta94b} 
Here the spectral weight exclusively concentrates at $k\approx\pi$ and 
finite but small $\omega\ll J$. 
We finally ask whether the gap in $S^{zz}(k,\omega)$
again closes  at the OH-AFM transition if $J_z/J$ is increased further.
Figure~\ref{Skw}(d) shows that the gap indeed closes, at $J_z/J=2.138(\simeq J_{z,c2}/J)$,
but this time at momentum $k=\pi$, reflecting the lattice-period
doubling in the AFM phase. 
Obviously, $S^{zz}(k,\omega)$ follows the behavior of the neutral gap 
$\Delta_{n}$ shown in Fig.~\ref{gaps}.
In the AFM phase [see Fig.~\ref{Skw}(e) with $J_z/J=3$],
the dispersion becomes flattened with a large excitation gap that opens 
at $k=\pi$, however. 
That is, the dynamical spin structure factor shows a distinct behavior 
in each phase of the spin-1 model with single-ion anisotropy. 
Interestingly, the results obtained in the EH, OH, and AFM phases are 
similar to those for the MI, HI, and DW phases of the 1D
EBHM.\cite{ELF14} This corroborates that the spin-1 model can be taken 
as an effective model for the EBHM with $n_b=2$.

\section{Summary}
\label{summary}
We studied the topological properties of the aniso\-tropic spin-1 $XXZ$ 
model with single-ion anisotropy 
in close analogy to a recent investigation of the extended Bose-Hubbard model (EBHM) 
with a nearest-neighbor repulsion.\cite{ELF14} The focus was on the 
nontrivial Haldane phase as well.  The phase boundaries between trivial
phases [even Haldane (EH) and AFM phases] and nontrivial odd Haldane (OH) phase
were determined numerically with high precision via the central
charge. The ground-state phase diagram resembles  those of  the restricted EBHM 
with a maximum number of bosons per site $n_b=2$, but the topological phase takes a much narrower  
region in the parameter space. Simulating the spin and neutral gaps, which correspond to
the single-particle respectively neutral gaps in the EBHM,  we confirmed the closing of the gap 
at the trivial-nontrivial quantum phase transition as for  the EBHM.

The degeneracy of the lowest entanglement level in the OH phase could be
observed by finite-system DMRG calculations with both  
periodic (P) and open (O) boundary conditions (BC). 
With PBC the lowest level in the entanglement spectrum
is fourfold degenerate in the OH phase; notably, the system-size dependence of the results is 
much stronger than for OBC. Adopting half spins ($S=1/2$) at the open 
edges, the twofold degeneracy corresponding to a single
artificial edge in the entanglement calculations can be detected easily.
This degeneracy will be lifted turning on a finite perturbation that breaks the inversion symmetry 
of the lattice, independently from the BC used.

We furthermore  used the dynamical DMRG technique to examine the dynamical 
spin structure $S^{zz}(k,\omega)$ which mimics the dynamical
density fluctuations in the EBHM. In the topological $S=1$ OH phase
a sinus-shaped dispersion was observed for finite anisotropy $D$ 
just as for the isotropic Heisenberg model and the Haldane insulator
state of the EBHM. Moreover, $S^{zz}(k,\omega)$ 
shows a significant different momentum and energy dependence in three different gapful phases 
for both the spin-1 model and the EBHM. We finally note that the
influence of the particle-hole symmetry-breaking term 
$\hat{\cal H}^{\prime}$ in Eq.~\eqref{eff-hamil-BH} on the properties of the constrained EBHM 
is almost negligible, not only for static but also for dynamical quantities.

\section*{Acknowledgments}
The authors would like to thank Y.~Fuji, F.~G{\"o}hmann, 
F.~Lange, S.~Nishimoto, and F.~Pollmann for valuable discussions. 
This work was supported by Deutsche Forschungsgemeinschaft (Germany),
SFB 652, B5.

\appendix

\section{Haldane gap}
\label{sec:Haldane-gap}

\begin{table}[tb]
 \begin{center}
  \begin{tabular}{c|c|c|c|c}
   \hline
   Method                  & $L$ & $\Delta$        & T         & BC\\ \hline \hline
   QMC\cite{TK01}          & 128 & 0.41048(6)      & 0.015625  & PBC\\
   DMRG\cite{WhiteHuse93}  & 120 & 0.41050(2)      & 0         & OBC\\
   ED\cite{NT09}           &  24 & 0.41047(8)      & 0         & TBC\\
   DMRG\cite{UK11}         & 2048& 0.4104792485(4) & 0         & OBC\\
   DMRG (this work)        &  96 & 0.4104792(7)    & 0         & PBC\\
   DMRG (this work)        & 128 & 0.41047924(4)   & 0         & PBC
  \end{tabular} 
  \caption{First excitation gap $\Delta$ in the spin-1 $XXZ$  chain 
  as obtained by  QMC, ED, and DMRG for a system size $L$, 
  at temperature $T$, using the specified boundary conditions BC.}
  \label{tab:Haldane-Gap}
 \end{center}
\end{table}

\begin{figure}[bt]
 \begin{center}
  \includegraphics[clip,width=0.9\columnwidth]{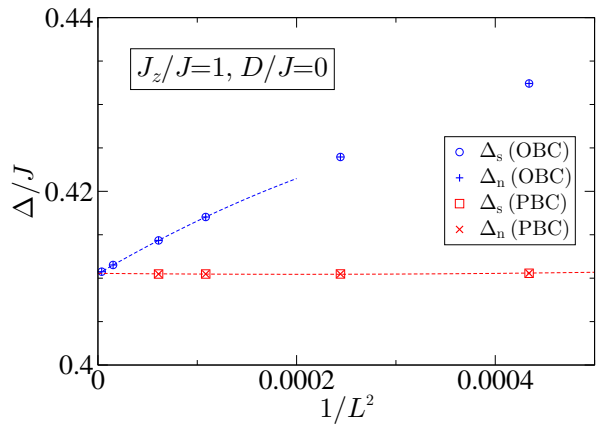}
  \end{center}
 \caption{(Color online) Finite-size scaling of the excitation gaps at the Heisenberg
 point ($D=0$ and $J_z/J=1$).
 }
\label{fig:Haldane-Gap}
\end{figure}

After Haldane's conjecture about the finite excitation gaps
for integer-spin chains,\cite{Ha83} it was a challenging issue 
to estimate these so-called Haldane gaps numerically  
(note that even the spin-1 $XXZ$ Heisenberg chain is not
integrable). White presented the first accurate 
DMRG results for the Haldane gap,\cite{Wh92}
and subsequently a series of more elaborated
DMRG,\cite{Wh93,WhiteHuse93,UK11} QMC~\cite{TK01} and
ED~\cite{NT09} studies 
have been performed. However, only OBC have
been used within the DMRG framework so far, mainly because of the smaller computational costs. 
In this Appendix, we demonstrate---at least for the spin-1 Heisenberg
model---that the Haldane gap can 
also be determined using PBC, and the system-size dependence of 
the gap is much smaller than those with OBC adopting
the half spin edges [cf. Fig.~\ref{Haldane-state}(b)].
Hence any finite-size scaling is needless.

Figure~\ref{fig:Haldane-Gap} presents the finite-size extrapolation 
of the corresponding spin and neutral excitation gaps, as defined in Sec.~\ref{sec:gap},
for both OBC and PBC. The spin and neutral gaps become equal ($\Delta$) only at 
the Heisenberg point for $D=0$ [cf. Fig.~\ref{gaps}(a)].
Computing $\Delta$ for systems with up to $L=512$ sites and OBC, we can extrapolate 
the results to the thermodynamic limit and obtain 
$\Delta=0.41050(3)$ (in agreement with Ref.~[\onlinecite{WhiteHuse93}]). 
On the other hand, the first excitation gaps $\Delta=0.41047924(4)$
obtained with PBC and up to $m=4800$ density-matrix states
show almost no finite-size dependence; see also the 
raw data for $L=96$ and $L=128$ in Table~\ref{tab:Haldane-Gap}.
This value is very close to the (low-temperature) QMC\cite{TK01} 
and ED\cite{NT09} results
and shows a perfect agreement with the very recent non-Abelian DMRG data
with OBC.\cite{UK11}
Let us emphasize that although the accessible system size is rather limited for PBC,
$\Delta$ for PBC is always lower than for OBC. Most notably, the system-size 
dependence is almost negligible (for enough large $L$),
so that sophisticated extrapolation techniques or 
the use of special boundary conditions,\cite{NT09,UK11}
are no longer mandatory for analyzing the Haldane gap in spin-1 chains.

\bibliographystyle{apsrev4-1}

\begin{thebibliography}{40}%
\makeatletter
\providecommand \@ifxundefined [1]{%
 \@ifx{#1\undefined}
}%
\providecommand \@ifnum [1]{%
 \ifnum #1\expandafter \@firstoftwo
 \else \expandafter \@secondoftwo
 \fi
}%
\providecommand \@ifx [1]{%
 \ifx #1\expandafter \@firstoftwo
 \else \expandafter \@secondoftwo
 \fi
}%
\providecommand \natexlab [1]{#1}%
\providecommand \enquote  [1]{``#1''}%
\providecommand \bibnamefont  [1]{#1}%
\providecommand \bibfnamefont [1]{#1}%
\providecommand \citenamefont [1]{#1}%
\providecommand \href@noop [0]{\@secondoftwo}%
\providecommand \href [0]{\begingroup \@sanitize@url \@href}%
\providecommand \@href[1]{\@@startlink{#1}\@@href}%
\providecommand \@@href[1]{\endgroup#1\@@endlink}%
\providecommand \@sanitize@url [0]{\catcode `\\12\catcode `\$12\catcode
  `\&12\catcode `\#12\catcode `\^12\catcode `\_12\catcode `\%12\relax}%
\providecommand \@@startlink[1]{}%
\providecommand \@@endlink[0]{}%
\providecommand \url  [0]{\begingroup\@sanitize@url \@url }%
\providecommand \@url [1]{\endgroup\@href {#1}{\urlprefix }}%
\providecommand \urlprefix  [0]{URL }%
\providecommand \Eprint [0]{\href }%
\providecommand \doibase [0]{http://dx.doi.org/}%
\providecommand \selectlanguage [0]{\@gobble}%
\providecommand \bibinfo  [0]{\@secondoftwo}%
\providecommand \bibfield  [0]{\@secondoftwo}%
\providecommand \translation [1]{[#1]}%
\providecommand \BibitemOpen [0]{}%
\providecommand \bibitemStop [0]{}%
\providecommand \bibitemNoStop [0]{.\EOS\space}%
\providecommand \EOS [0]{\spacefactor3000\relax}%
\providecommand \BibitemShut  [1]{\csname bibitem#1\endcsname}%
\let\auto@bib@innerbib\@empty
\bibitem [{\citenamefont {Haldane}(1983)}]{Ha83}%
  \BibitemOpen
  \bibfield  {author} {\bibinfo {author} {\bibfnamefont {F.~D.~M.}\
  \bibnamefont {Haldane}},\ }\href {\doibase 10.1103/PhysRevLett.50.1153}
  {\bibfield  {journal} {\bibinfo  {journal} {Phys. Rev. Lett.}\ }\textbf
  {\bibinfo {volume} {50}},\ \bibinfo {pages} {1153} (\bibinfo {year}
  {1983})}\BibitemShut {NoStop}%
\bibitem [{\citenamefont {Buyers}\ \emph {et~al.}(1986)\citenamefont {Buyers},
  \citenamefont {Morra}, \citenamefont {Armstrong}, \citenamefont {Hogan},
  \citenamefont {Gerlach},\ and\ \citenamefont {Hirakawa}}]{BMAHGH86}%
  \BibitemOpen
  \bibfield  {author} {\bibinfo {author} {\bibfnamefont {W.~J.~L.}\
  \bibnamefont {Buyers}}, \bibinfo {author} {\bibfnamefont {R.~M.}\
  \bibnamefont {Morra}}, \bibinfo {author} {\bibfnamefont {R.~L.}\ \bibnamefont
  {Armstrong}}, \bibinfo {author} {\bibfnamefont {M.~J.}\ \bibnamefont
  {Hogan}}, \bibinfo {author} {\bibfnamefont {P.}~\bibnamefont {Gerlach}}, \
  and\ \bibinfo {author} {\bibfnamefont {K.}~\bibnamefont {Hirakawa}},\ }\href
  {\doibase 10.1103/PhysRevLett.56.371} {\bibfield  {journal} {\bibinfo
  {journal} {Phys. Rev. Lett.}\ }\textbf {\bibinfo {volume} {56}},\ \bibinfo
  {pages} {371} (\bibinfo {year} {1986})}\BibitemShut {NoStop}%
\bibitem [{\citenamefont {Renard}\ \emph {et~al.}(1987)\citenamefont {Renard},
  \citenamefont {Verdaguer}, \citenamefont {Regnault}, \citenamefont
  {Erkelens}, \citenamefont {Rossat-Mignod},\ and\ \citenamefont
  {Stirling}}]{RVRERS87}%
  \BibitemOpen
  \bibfield  {author} {\bibinfo {author} {\bibfnamefont {J.~P.}\ \bibnamefont
  {Renard}}, \bibinfo {author} {\bibfnamefont {M.}~\bibnamefont {Verdaguer}},
  \bibinfo {author} {\bibfnamefont {L.~P.}\ \bibnamefont {Regnault}}, \bibinfo
  {author} {\bibfnamefont {W.~A.~C.}\ \bibnamefont {Erkelens}}, \bibinfo
  {author} {\bibfnamefont {J.}~\bibnamefont {Rossat-Mignod}}, \ and\ \bibinfo
  {author} {\bibfnamefont {W.~G.}\ \bibnamefont {Stirling}},\ }\href {\doibase
  10.1209/0295-5075/3/8/013} {\bibfield  {journal} {\bibinfo  {journal}
  {Europhys. Lett.}\ }\textbf {\bibinfo {volume} {3}},\ \bibinfo {pages} {945}
  (\bibinfo {year} {1987})}\BibitemShut {NoStop}%
\bibitem [{\citenamefont {Ma}\ \emph {et~al.}(1992)\citenamefont {Ma},
  \citenamefont {Broholm}, \citenamefont {Reich}, \citenamefont {Sternlieb},\
  and\ \citenamefont {Erwin}}]{MBRSE92}%
  \BibitemOpen
  \bibfield  {author} {\bibinfo {author} {\bibfnamefont {S.}~\bibnamefont
  {Ma}}, \bibinfo {author} {\bibfnamefont {C.}~\bibnamefont {Broholm}},
  \bibinfo {author} {\bibfnamefont {D.~H.}\ \bibnamefont {Reich}}, \bibinfo
  {author} {\bibfnamefont {B.~J.}\ \bibnamefont {Sternlieb}}, \ and\ \bibinfo
  {author} {\bibfnamefont {R.~W.}\ \bibnamefont {Erwin}},\ }\href {\doibase
  10.1103/PhysRevLett.69.3571} {\bibfield  {journal} {\bibinfo  {journal}
  {Phys. Rev. Lett.}\ }\textbf {\bibinfo {volume} {69}},\ \bibinfo {pages}
  {3571} (\bibinfo {year} {1992})}\BibitemShut {NoStop}%
\bibitem [{\citenamefont {Affleck}\ \emph {et~al.}(1987)\citenamefont
  {Affleck}, \citenamefont {Kennedy}, \citenamefont {Lieb},\ and\ \citenamefont
  {Tasaki}}]{AKLT87}%
  \BibitemOpen
  \bibfield  {author} {\bibinfo {author} {\bibfnamefont {I.}~\bibnamefont
  {Affleck}}, \bibinfo {author} {\bibfnamefont {T.}~\bibnamefont {Kennedy}},
  \bibinfo {author} {\bibfnamefont {E.~H.}\ \bibnamefont {Lieb}}, \ and\
  \bibinfo {author} {\bibfnamefont {H.}~\bibnamefont {Tasaki}},\ }\href
  {\doibase 10.1103/PhysRevLett.59.799} {\bibfield  {journal} {\bibinfo
  {journal} {Phys. Rev. Lett.}\ }\textbf {\bibinfo {volume} {59}},\ \bibinfo
  {pages} {799} (\bibinfo {year} {1987})}\BibitemShut {NoStop}%
\bibitem [{\citenamefont {Kennedy}\ and\ \citenamefont {Tasaki}(1992)}]{KT92}%
  \BibitemOpen
  \bibfield  {author} {\bibinfo {author} {\bibfnamefont {T.}~\bibnamefont
  {Kennedy}}\ and\ \bibinfo {author} {\bibfnamefont {H.}~\bibnamefont
  {Tasaki}},\ }\href {\doibase 10.1103/PhysRevB.45.304} {\bibfield  {journal}
  {\bibinfo  {journal} {Phys. Rev. B}\ }\textbf {\bibinfo {volume} {45}},\
  \bibinfo {pages} {304} (\bibinfo {year} {1992})}\BibitemShut {NoStop}%
\bibitem [{\citenamefont {Chen}\ \emph {et~al.}(2003)\citenamefont {Chen},
  \citenamefont {Hida},\ and\ \citenamefont {Sanctuary}}]{CHS03}%
  \BibitemOpen
  \bibfield  {author} {\bibinfo {author} {\bibfnamefont {W.}~\bibnamefont
  {Chen}}, \bibinfo {author} {\bibfnamefont {K.}~\bibnamefont {Hida}}, \ and\
  \bibinfo {author} {\bibfnamefont {B.~C.}\ \bibnamefont {Sanctuary}},\ }\href
  {\doibase 10.1103/PhysRevB.67.104401} {\bibfield  {journal} {\bibinfo
  {journal} {Phys. Rev. B}\ }\textbf {\bibinfo {volume} {67}},\ \bibinfo
  {pages} {104401} (\bibinfo {year} {2003})}\BibitemShut {NoStop}%
\bibitem [{\citenamefont {Nomura}(1995)}]{No95}%
  \BibitemOpen
  \bibfield  {author} {\bibinfo {author} {\bibfnamefont {K.}~\bibnamefont
  {Nomura}},\ }\href {\doibase 10.1088/0305-4470/28/19/003} {\bibfield
  {journal} {\bibinfo  {journal} {J. Phys. A}\ }\textbf {\bibinfo {volume}
  {28}},\ \bibinfo {pages} {5451} (\bibinfo {year} {1995})}\BibitemShut
  {NoStop}%
\bibitem [{\citenamefont {Gu}\ and\ \citenamefont {Wen}(2009)}]{GW09}%
  \BibitemOpen
  \bibfield  {author} {\bibinfo {author} {\bibfnamefont {Z.-C.}\ \bibnamefont
  {Gu}}\ and\ \bibinfo {author} {\bibfnamefont {X.-G.}\ \bibnamefont {Wen}},\
  }\href {\doibase 10.1103/PhysRevB.80.155131} {\bibfield  {journal} {\bibinfo
  {journal} {Phys. Rev. B}\ }\textbf {\bibinfo {volume} {80}},\ \bibinfo
  {pages} {155131} (\bibinfo {year} {2009})}\BibitemShut {NoStop}%
\bibitem [{\citenamefont {Pollmann}\ \emph {et~al.}(2010)\citenamefont
  {Pollmann}, \citenamefont {Turner}, \citenamefont {Berg},\ and\ \citenamefont
  {Oshikawa}}]{PTBO10}%
  \BibitemOpen
  \bibfield  {author} {\bibinfo {author} {\bibfnamefont {F.}~\bibnamefont
  {Pollmann}}, \bibinfo {author} {\bibfnamefont {A.~M.}\ \bibnamefont
  {Turner}}, \bibinfo {author} {\bibfnamefont {E.}~\bibnamefont {Berg}}, \ and\
  \bibinfo {author} {\bibfnamefont {M.}~\bibnamefont {Oshikawa}},\ }\href
  {\doibase 10.1103/PhysRevB.81.064439} {\bibfield  {journal} {\bibinfo
  {journal} {Phys. Rev. B}\ }\textbf {\bibinfo {volume} {81}},\ \bibinfo
  {pages} {064439} (\bibinfo {year} {2010})}\BibitemShut {NoStop}%
\bibitem [{\citenamefont {Tonegawa}\ \emph {et~al.}(2011)\citenamefont
  {Tonegawa}, \citenamefont {Okamoto}, \citenamefont {Nakano}, \citenamefont
  {Sakai}, \citenamefont {Nomura},\ and\ \citenamefont {Kaburagi}}]{TONSNK11}%
  \BibitemOpen
  \bibfield  {author} {\bibinfo {author} {\bibfnamefont {T.}~\bibnamefont
  {Tonegawa}}, \bibinfo {author} {\bibfnamefont {K.}~\bibnamefont {Okamoto}},
  \bibinfo {author} {\bibfnamefont {H.}~\bibnamefont {Nakano}}, \bibinfo
  {author} {\bibfnamefont {T.}~\bibnamefont {Sakai}}, \bibinfo {author}
  {\bibfnamefont {K.}~\bibnamefont {Nomura}}, \ and\ \bibinfo {author}
  {\bibfnamefont {M.}~\bibnamefont {Kaburagi}},\ }\href {\doibase
  10.1143/JPSJ.80.043001} {\bibfield  {journal} {\bibinfo  {journal} {J. Phys.
  Soc. Jpn.}\ }\textbf {\bibinfo {volume} {80}},\ \bibinfo {pages} {043001}
  (\bibinfo {year} {2011})}\BibitemShut {NoStop}%
\bibitem [{\citenamefont {Okamoto}\ \emph
  {et~al.}(2011{\natexlab{a}})\citenamefont {Okamoto}, \citenamefont
  {Tonegawa}, \citenamefont {Nakano}, \citenamefont {Sakai}, \citenamefont
  {Nomura},\ and\ \citenamefont {Kaburagi}}]{OTNSNK11}%
  \BibitemOpen
  \bibfield  {author} {\bibinfo {author} {\bibfnamefont {K.}~\bibnamefont
  {Okamoto}}, \bibinfo {author} {\bibfnamefont {T.}~\bibnamefont {Tonegawa}},
  \bibinfo {author} {\bibfnamefont {H.}~\bibnamefont {Nakano}}, \bibinfo
  {author} {\bibfnamefont {T.}~\bibnamefont {Sakai}}, \bibinfo {author}
  {\bibfnamefont {K.}~\bibnamefont {Nomura}}, \ and\ \bibinfo {author}
  {\bibfnamefont {M.}~\bibnamefont {Kaburagi}},\ }\href {\doibase
  10.1088/1742-6596/302/1/012014} {\bibfield  {journal} {\bibinfo  {journal}
  {J. Phys.: Conf. Ser.}\ }\textbf {\bibinfo {volume} {302}},\ \bibinfo {pages}
  {012014} (\bibinfo {year} {2011}{\natexlab{a}})}\BibitemShut {NoStop}%
\bibitem [{\citenamefont {Okamoto}\ \emph
  {et~al.}(2011{\natexlab{b}})\citenamefont {Okamoto}, \citenamefont
  {Tonegawa}, \citenamefont {Nakano}, \citenamefont {Sakai}, \citenamefont
  {Nomura},\ and\ \citenamefont {Kaburagi}}]{OTNSNK11b}%
  \BibitemOpen
  \bibfield  {author} {\bibinfo {author} {\bibfnamefont {K.}~\bibnamefont
  {Okamoto}}, \bibinfo {author} {\bibfnamefont {T.}~\bibnamefont {Tonegawa}},
  \bibinfo {author} {\bibfnamefont {H.}~\bibnamefont {Nakano}}, \bibinfo
  {author} {\bibfnamefont {T.}~\bibnamefont {Sakai}}, \bibinfo {author}
  {\bibfnamefont {K.}~\bibnamefont {Nomura}}, \ and\ \bibinfo {author}
  {\bibfnamefont {M.}~\bibnamefont {Kaburagi}},\ }\href {\doibase
  10.1088/1742-6596/320/1/012018} {\bibfield  {journal} {\bibinfo  {journal}
  {J. Phys.: Conf. Ser.}\ }\textbf {\bibinfo {volume} {320}},\ \bibinfo {pages}
  {012018} (\bibinfo {year} {2011}{\natexlab{b}})}\BibitemShut {NoStop}%
\bibitem [{\citenamefont {Tzeng}(2012)}]{Tz12}%
  \BibitemOpen
  \bibfield  {author} {\bibinfo {author} {\bibfnamefont {Y.-C.}\ \bibnamefont
  {Tzeng}},\ }\href {\doibase 10.1103/PhysRevB.86.024403} {\bibfield  {journal}
  {\bibinfo  {journal} {Phys. Rev. B}\ }\textbf {\bibinfo {volume} {86}},\
  \bibinfo {pages} {024403} (\bibinfo {year} {2012})}\BibitemShut {NoStop}%
\bibitem [{\citenamefont {Pollmann}\ \emph {et~al.}(2012)\citenamefont
  {Pollmann}, \citenamefont {Berg}, \citenamefont {Turner},\ and\ \citenamefont
  {Oshikawa}}]{PBTO12}%
  \BibitemOpen
  \bibfield  {author} {\bibinfo {author} {\bibfnamefont {F.}~\bibnamefont
  {Pollmann}}, \bibinfo {author} {\bibfnamefont {E.}~\bibnamefont {Berg}},
  \bibinfo {author} {\bibfnamefont {A.~M.}\ \bibnamefont {Turner}}, \ and\
  \bibinfo {author} {\bibfnamefont {M.}~\bibnamefont {Oshikawa}},\ }\href
  {\doibase 10.1103/PhysRevB.85.075125} {\bibfield  {journal} {\bibinfo
  {journal} {Phys. Rev. B}\ }\textbf {\bibinfo {volume} {85}},\ \bibinfo
  {pages} {075125} (\bibinfo {year} {2012})}\BibitemShut {NoStop}%
\bibitem [{\citenamefont {Kj\"all}\ \emph {et~al.}(2013)\citenamefont
  {Kj\"all}, \citenamefont {Zaletel}, \citenamefont {Mong}, \citenamefont
  {Bardarson},\ and\ \citenamefont {Pollmann}}]{KZMBP13}%
  \BibitemOpen
  \bibfield  {author} {\bibinfo {author} {\bibfnamefont {J.~A.}\ \bibnamefont
  {Kj\"all}}, \bibinfo {author} {\bibfnamefont {M.~P.}\ \bibnamefont
  {Zaletel}}, \bibinfo {author} {\bibfnamefont {R.~S.~K.}\ \bibnamefont
  {Mong}}, \bibinfo {author} {\bibfnamefont {J.~H.}\ \bibnamefont {Bardarson}},
  \ and\ \bibinfo {author} {\bibfnamefont {F.}~\bibnamefont {Pollmann}},\
  }\href {\doibase 10.1103/PhysRevB.87.235106} {\bibfield  {journal} {\bibinfo
  {journal} {Phys. Rev. B}\ }\textbf {\bibinfo {volume} {87}},\ \bibinfo
  {pages} {235106} (\bibinfo {year} {2013})}\BibitemShut {NoStop}%
\bibitem [{\citenamefont {{Dalla Torre}}\ \emph {et~al.}(2006)\citenamefont
  {{Dalla Torre}}, \citenamefont {Berg},\ and\ \citenamefont {Altman}}]{DBA06}%
  \BibitemOpen
  \bibfield  {author} {\bibinfo {author} {\bibfnamefont {E.~G.}\ \bibnamefont
  {{Dalla Torre}}}, \bibinfo {author} {\bibfnamefont {E.}~\bibnamefont {Berg}},
  \ and\ \bibinfo {author} {\bibfnamefont {E.}~\bibnamefont {Altman}},\ }\href
  {\doibase 10.1103/PhysRevLett.97.260401} {\bibfield  {journal} {\bibinfo
  {journal} {Phys. Rev. Lett.}\ }\textbf {\bibinfo {volume} {97}},\ \bibinfo
  {pages} {260401} (\bibinfo {year} {2006})}\BibitemShut {NoStop}%
\bibitem [{\citenamefont {Ejima}\ \emph {et~al.}(2014)\citenamefont {Ejima},
  \citenamefont {Lange},\ and\ \citenamefont {Fehske}}]{ELF14}%
  \BibitemOpen
  \bibfield  {author} {\bibinfo {author} {\bibfnamefont {S.}~\bibnamefont
  {Ejima}}, \bibinfo {author} {\bibfnamefont {F.}~\bibnamefont {Lange}}, \ and\
  \bibinfo {author} {\bibfnamefont {H.}~\bibnamefont {Fehske}},\ }\href
  {\doibase 10.1103/PhysRevLett.113.020401} {\bibfield  {journal} {\bibinfo
  {journal} {Phys. Rev. Lett.}\ }\textbf {\bibinfo {volume} {113}},\ \bibinfo
  {pages} {020401} (\bibinfo {year} {2014})}\BibitemShut {NoStop}%
\bibitem [{\citenamefont {Berg}\ \emph {et~al.}(2008)\citenamefont {Berg},
  \citenamefont {{Dalla Torre}}, \citenamefont {Giamarchi},\ and\ \citenamefont
  {Altman}}]{BDGA08}%
  \BibitemOpen
  \bibfield  {author} {\bibinfo {author} {\bibfnamefont {E.}~\bibnamefont
  {Berg}}, \bibinfo {author} {\bibfnamefont {E.~G.}~\bibnamefont {{Dalla Torre}}},
  \bibinfo {author} {\bibfnamefont {T.}~\bibnamefont {Giamarchi}}, \ and\
  \bibinfo {author} {\bibfnamefont {E.}~\bibnamefont {Altman}},\ }\href
  {\doibase 10.1103/PhysRevB.77.245119} {\bibfield  {journal} {\bibinfo
  {journal} {Phys. Rev. B}\ }\textbf {\bibinfo {volume} {77}},\ \bibinfo
  {pages} {245119} (\bibinfo {year} {2008})}\BibitemShut {NoStop}%
\bibitem [{\citenamefont {White}(1992)}]{Wh92}%
  \BibitemOpen
  \bibfield  {author} {\bibinfo {author} {\bibfnamefont {S.~R.}\ \bibnamefont
  {White}},\ }\href@noop {} {\bibfield  {journal} {\bibinfo  {journal} {Phys.
  Rev. Lett.}\ }\textbf {\bibinfo {volume} {69}},\ \bibinfo {pages} {2863}
  (\bibinfo {year} {1992})}\BibitemShut {NoStop}%
\bibitem [{\citenamefont {White}(1993)}]{Wh93}%
  \BibitemOpen
  \bibfield  {author} {\bibinfo {author} {\bibfnamefont {S.~R.}\ \bibnamefont
  {White}},\ }\href@noop {} {\bibfield  {journal} {\bibinfo  {journal} {Phys.
  Rev. B}\ }\textbf {\bibinfo {volume} {48}},\ \bibinfo {pages} {10345}
  (\bibinfo {year} {1993})}\BibitemShut {NoStop}%
\bibitem [{\citenamefont {Jeckelmann}\ and\ \citenamefont
  {Fehske}(2007)}]{JF07}%
  \BibitemOpen
  \bibfield  {author} {\bibinfo {author} {\bibfnamefont {E.}~\bibnamefont
  {Jeckelmann}}\ and\ \bibinfo {author} {\bibfnamefont {H.}~\bibnamefont
  {Fehske}},\ }\href@noop {} {\bibfield  {journal} {\bibinfo  {journal}
  {Rivista del Nuovo Cimento}\ }\textbf {\bibinfo {volume} {30}},\ \bibinfo
  {pages} {259} (\bibinfo {year} {2007})}\BibitemShut {NoStop}%
\bibitem [{\citenamefont {Torre}(2013)}]{Da13}%
  \BibitemOpen
  \bibfield  {author} {\bibinfo {author} {\bibfnamefont {E.~G.~D.}\
  \bibnamefont {Torre}},\ }\href {\doibase 10.1088/0953-4075/46/8/085303}
  {\bibfield  {journal} {\bibinfo  {journal} {J. Phys. B: At. Mol. Opt. Phys.}\
  }\textbf {\bibinfo {volume} {46}},\ \bibinfo {pages} {085303} (\bibinfo
  {year} {2013})}\BibitemShut {NoStop}%
\bibitem [{\citenamefont {Jeckelmann}(2002)}]{Je02b}%
  \BibitemOpen
  \bibfield  {author} {\bibinfo {author} {\bibfnamefont {E.}~\bibnamefont
  {Jeckelmann}},\ }\href@noop {} {\bibfield  {journal} {\bibinfo  {journal}
  {Phys. Rev. B}\ }\textbf {\bibinfo {volume} {66}},\ \bibinfo {pages} {045114}
  (\bibinfo {year} {2002})}\BibitemShut {NoStop}%
\bibitem [{\citenamefont {Schulz}(1986)}]{Sc86}%
  \BibitemOpen
  \bibfield  {author} {\bibinfo {author} {\bibfnamefont {H.~J.}\ \bibnamefont
  {Schulz}},\ }\href {\doibase 10.1103/PhysRevB.34.6372} {\bibfield  {journal}
  {\bibinfo  {journal} {Phys. Rev. B}\ }\textbf {\bibinfo {volume} {34}},\
  \bibinfo {pages} {6372} (\bibinfo {year} {1986})}\BibitemShut {NoStop}%
\bibitem [{\citenamefont {den Nijs}\ and\ \citenamefont
  {Rommelse}(1989)}]{NR89}%
  \BibitemOpen
  \bibfield  {author} {\bibinfo {author} {\bibfnamefont {M.}~\bibnamefont {den
  Nijs}}\ and\ \bibinfo {author} {\bibfnamefont {K.}~\bibnamefont {Rommelse}},\
  }\href {\doibase 10.1103/PhysRevB.40.4709} {\bibfield  {journal} {\bibinfo
  {journal} {Phys. Rev. B}\ }\textbf {\bibinfo {volume} {40}},\ \bibinfo
  {pages} {4709} (\bibinfo {year} {1989})}\BibitemShut {NoStop}%
\bibitem [{\citenamefont {Li}\ and\ \citenamefont {Haldane}(2008)}]{LH08}%
  \BibitemOpen
  \bibfield  {author} {\bibinfo {author} {\bibfnamefont {H.}~\bibnamefont
  {Li}}\ and\ \bibinfo {author} {\bibfnamefont {F.~D.~M.}\ \bibnamefont
  {Haldane}},\ }\href {\doibase 10.1103/PhysRevLett.101.010504} {\bibfield
  {journal} {\bibinfo  {journal} {Phys. Rev. Lett.}\ }\textbf {\bibinfo
  {volume} {101}},\ \bibinfo {pages} {010504} (\bibinfo {year}
  {2008})}\BibitemShut {NoStop}%
\bibitem [{\citenamefont {Calabrese}\ and\ \citenamefont {Cardy}()}]{CC04}%
  \BibitemOpen
  \bibfield  {author} {\bibinfo {author} {\bibfnamefont {P.}~\bibnamefont
  {Calabrese}}\ and\ \bibinfo {author} {\bibfnamefont {J.}~\bibnamefont
  {Cardy}},\ }\href@noop {} {\bibfield  {journal} {\bibinfo  {journal} {J.
  Stat. Mech.}\ }\textbf {\bibinfo {volume} {(2004)}},\ \bibinfo {pages}
  {P06002}}\BibitemShut {NoStop}%
\bibitem [{\citenamefont {Nishimoto}(2011)}]{Ni11}%
  \BibitemOpen
  \bibfield  {author} {\bibinfo {author} {\bibfnamefont {S.}~\bibnamefont
  {Nishimoto}},\ }\href@noop {} {\bibfield  {journal} {\bibinfo  {journal}
  {Phys. Rev. B}\ }\textbf {\bibinfo {volume} {84}},\ \bibinfo {pages} {195108}
  (\bibinfo {year} {2011})}\BibitemShut {NoStop}%
\bibitem [{\citenamefont {Takahashi}(1994)}]{Ta94b}%
  \BibitemOpen
  \bibfield  {author} {\bibinfo {author} {\bibfnamefont {M.}~\bibnamefont
  {Takahashi}},\ }\href {\doibase 10.1103/PhysRevB.50.3045} {\bibfield
  {journal} {\bibinfo  {journal} {Phys. Rev. B}\ }\textbf {\bibinfo {volume}
  {50}},\ \bibinfo {pages} {3045} (\bibinfo {year} {1994})}\BibitemShut
  {NoStop}%
\bibitem [{\citenamefont {White}\ and\ \citenamefont {Affleck}(2008)}]{WA08}%
  \BibitemOpen
  \bibfield  {author} {\bibinfo {author} {\bibfnamefont {S.~R.}\ \bibnamefont
  {White}}\ and\ \bibinfo {author} {\bibfnamefont {I.}~\bibnamefont
  {Affleck}},\ }\href {\doibase 10.1103/PhysRevB.77.134437} {\bibfield
  {journal} {\bibinfo  {journal} {Phys. Rev. B}\ }\textbf {\bibinfo {volume}
  {77}},\ \bibinfo {pages} {134437} (\bibinfo {year} {2008})}\BibitemShut
  {NoStop}%
\bibitem [{\citenamefont {Kitazawa}\ \emph {et~al.}(1996)\citenamefont
  {Kitazawa}, \citenamefont {Nomura},\ and\ \citenamefont {Okamoto}}]{KNO96}%
  \BibitemOpen
  \bibfield  {author} {\bibinfo {author} {\bibfnamefont {A.}~\bibnamefont
  {Kitazawa}}, \bibinfo {author} {\bibfnamefont {K.}~\bibnamefont {Nomura}}, \
  and\ \bibinfo {author} {\bibfnamefont {K.}~\bibnamefont {Okamoto}},\ }\href
  {\doibase 10.1103/PhysRevLett.76.4038} {\bibfield  {journal} {\bibinfo
  {journal} {Phys. Rev. Lett.}\ }\textbf {\bibinfo {volume} {76}},\ \bibinfo
  {pages} {4038} (\bibinfo {year} {1996})}\BibitemShut {NoStop}%
\bibitem [{\citenamefont {Liu}\ \emph {et~al.}(2014)\citenamefont {Liu},
  \citenamefont {Li}, \citenamefont {You}, \citenamefont {Su},\ and\
  \citenamefont {Tian}}]{LLYST14}%
  \BibitemOpen
  \bibfield  {author} {\bibinfo {author} {\bibfnamefont {G.-H.}\ \bibnamefont
  {Liu}}, \bibinfo {author} {\bibfnamefont {W.}~\bibnamefont {Li}}, \bibinfo
  {author} {\bibfnamefont {W.-L.}\ \bibnamefont {You}}, \bibinfo {author}
  {\bibfnamefont {G.}~\bibnamefont {Su}}, \ and\ \bibinfo {author}
  {\bibfnamefont {G.-S.}\ \bibnamefont {Tian}},\ }\href {\doibase
  http://dx.doi.org/10.1016/j.physb.2014.03.007} {\bibfield  {journal}
  {\bibinfo  {journal} {Physica B}\ }\textbf {\bibinfo {volume} {443}},\
  \bibinfo {pages} {63 } (\bibinfo {year} {2014})}\BibitemShut {NoStop}%
\bibitem [{\citenamefont {Ueda}\ \emph {et~al.}(2008)\citenamefont {Ueda},
  \citenamefont {Nakano},\ and\ \citenamefont {Kusakabe}}]{UNK08}%
  \BibitemOpen
  \bibfield  {author} {\bibinfo {author} {\bibfnamefont {H.}~\bibnamefont
  {Ueda}}, \bibinfo {author} {\bibfnamefont {H.}~\bibnamefont {Nakano}}, \ and\
  \bibinfo {author} {\bibfnamefont {K.}~\bibnamefont {Kusakabe}},\ }\href
  {\doibase 10.1103/PhysRevB.78.224402} {\bibfield  {journal} {\bibinfo
  {journal} {Phys. Rev. B}\ }\textbf {\bibinfo {volume} {78}},\ \bibinfo
  {pages} {224402} (\bibinfo {year} {2008})}\BibitemShut {NoStop}%
\bibitem [{\citenamefont {Kitazawa}\ and\ \citenamefont
  {Nomura}(1997{\natexlab{a}})}]{KN97a}%
  \BibitemOpen
  \bibfield  {author} {\bibinfo {author} {\bibfnamefont {A.}~\bibnamefont
  {Kitazawa}}\ and\ \bibinfo {author} {\bibfnamefont {K.}~\bibnamefont
  {Nomura}},\ }\href {\doibase 10.1143/JPSJ.66.3379} {\bibfield  {journal}
  {\bibinfo  {journal} {J. Phys. Soc. Jpn.}\ }\textbf {\bibinfo {volume}
  {66}},\ \bibinfo {pages} {3379} (\bibinfo {year}
  {1997}{\natexlab{a}})}\BibitemShut {NoStop}%
\bibitem [{\citenamefont {Kitazawa}\ and\ \citenamefont
  {Nomura}(1997{\natexlab{b}})}]{KN97b}%
  \BibitemOpen
  \bibfield  {author} {\bibinfo {author} {\bibfnamefont {A.}~\bibnamefont
  {Kitazawa}}\ and\ \bibinfo {author} {\bibfnamefont {K.}~\bibnamefont
  {Nomura}},\ }\href {\doibase 10.1143/JPSJ.66.3944} {\bibfield  {journal}
  {\bibinfo  {journal} {J. Phys. Soc. Jpn.}\ }\textbf {\bibinfo {volume}
  {66}},\ \bibinfo {pages} {3944} (\bibinfo {year}
  {1997}{\natexlab{b}})}\BibitemShut {NoStop}%
\bibitem [{\citenamefont {Todo}\ and\ \citenamefont {Kato}(2001)}]{TK01}%
  \BibitemOpen
  \bibfield  {author} {\bibinfo {author} {\bibfnamefont {S.}~\bibnamefont
  {Todo}}\ and\ \bibinfo {author} {\bibfnamefont {K.}~\bibnamefont {Kato}},\
  }\href {\doibase 10.1103/PhysRevLett.87.047203} {\bibfield  {journal}
  {\bibinfo  {journal} {Phys. Rev. Lett.}\ }\textbf {\bibinfo {volume} {87}},\
  \bibinfo {pages} {047203} (\bibinfo {year} {2001})}\BibitemShut {NoStop}%
\bibitem [{\citenamefont {White}\ and\ \citenamefont
  {Huse}(1993)}]{WhiteHuse93}%
  \BibitemOpen
  \bibfield  {author} {\bibinfo {author} {\bibfnamefont {S.~R.}\ \bibnamefont
  {White}}\ and\ \bibinfo {author} {\bibfnamefont {D.~A.}\ \bibnamefont
  {Huse}},\ }\href {\doibase 10.1103/PhysRevB.48.3844} {\bibfield  {journal}
  {\bibinfo  {journal} {Phys. Rev. B}\ }\textbf {\bibinfo {volume} {48}},\
  \bibinfo {pages} {3844} (\bibinfo {year} {1993})}\BibitemShut {NoStop}%
\bibitem [{\citenamefont {Nakano}\ and\ \citenamefont {Terai}(2009)}]{NT09}%
  \BibitemOpen
  \bibfield  {author} {\bibinfo {author} {\bibfnamefont {H.}~\bibnamefont
  {Nakano}}\ and\ \bibinfo {author} {\bibfnamefont {A.}~\bibnamefont {Terai}},\
  }\href {\doibase 10.1143/JPSJ.78.014003} {\bibfield  {journal} {\bibinfo
  {journal} {J. Phys. Soc. Jpn.}\ }\textbf {\bibinfo {volume} {78}},\ \bibinfo
  {pages} {014003} (\bibinfo {year} {2009})}\BibitemShut {NoStop}%
\bibitem [{\citenamefont {Ueda}\ and\ \citenamefont {Kusakabe}(2011)}]{UK11}%
  \BibitemOpen
  \bibfield  {author} {\bibinfo {author} {\bibfnamefont {H.}~\bibnamefont
  {Ueda}}\ and\ \bibinfo {author} {\bibfnamefont {K.}~\bibnamefont
  {Kusakabe}},\ }\href {\doibase 10.1103/PhysRevB.84.054446} {\bibfield
  {journal} {\bibinfo  {journal} {Phys. Rev. B}\ }\textbf {\bibinfo {volume}
  {84}},\ \bibinfo {pages} {054446} (\bibinfo {year} {2011})}\BibitemShut
  {NoStop}%
\end{thebibliography}

\end{document}